\documentclass[twocolumn]{aa}

\usepackage{graphicx}
\usepackage{txfonts}
\usepackage{lipsum}
\usepackage{subcaption}         
\usepackage{lscape}             
\usepackage{placeins}           
\usepackage{booktabs}                                

\begin{document}


     \title{Simultaneous radio, optical and X-ray monitoring of hard X-ray selected AGN: a variability study}

   \titlerunning{Multi-wavelength variability of hard X-ray selected AGN}

%

   \author{L. Hernández-García\inst{1,2}\fnmsep\thanks{Corresponding author: lorena.hernandez@mail.udp.cl}
        \and F. Panessa\inst{3}
        \and D. Williams-Baldwin\inst{4}
        \and P. Arévalo\inst{5,6}
        \and A. M. Muñoz Arancibia\inst{7,8}
        }

   \institute{Instituto de Estudios Astrof\'isicos, Facultad de Ingenier\'ia y Ciencias, Universidad Diego Portales, Av. Ej\'ercito Libertador 441, Santiago, Chile
  \and Centro Interdisciplinario de Data Science, Facultad de Ingenier\'ia y Ciencias, Universidad Diego Portales, Av. Ej\'ercito Libertador 441, Santiago, Chile
  \and INAF -- Istituto di Astrofisica e Planetologia Spaziali,
via del Fosso del Cavaliere 100, Roma, 00133, Italy
\and Jodrell Bank Centre for Astrophysics, School of Physics and Astronomy, The University of Manchester, Manchester, M13 9PL, UK
  \and Instituto de Física y Astronomía, Universidad de Valparaíso, Gran Bretaña 1111, Valparaíso, Chile
  \and Millennium Nucleus on Transversal Research and Technology to Explore Supermassive Black Holes (TITANS), Chile
  \and Millennium Institute of Astrophysics, Nuncio Monseñor Sotero Sanz 100, Of. 104, Providencia, Santiago, Chile
  \and Center for Mathematical Modeling, Universidad de Chile, Beauchef 851, Santiago 8370456, Chile
  }

   \date{Received September 30, 20XX}

 
  \abstract
   {AGN emission is intrinsically variable across the electromagnetic spectrum. Mapping the coupling between the accretion disk, the X-ray corona, and ejection flows is key to understanding the energy flow within the central engine.}
   {We aim to characterize the multi-wavelength variability of hard X-ray selected AGN across the radio, optical, and X-ray bands. Our goal is to determine the degree of coupling between these frequencies and to explore how variability features relate to the physical properties of the central engine, with special emphasis on the radio band.} 
   {The sample consists of 14 AGN selected from the \textit{INTEGRAL}/IBIS catalog. We analyzed multi-epoch observations from the Arcminute Microkelvin Imager (AMI-LA) at 15~GHz, the Zwicky Transient Facility (ZTF) in the $g$ and $r$ bands, and \textit{Swift}/XRT, covering a two-year monitoring period (2018--2020). Variability was quantified using the normalized excess variance, the fractional variability amplitude, and the Mexican Hat filter at 70- and 200-day timescales. Additionally, we characterized the radio-loudness of the sample using both X-ray and optical definitions, and evaluated the impact of variability on the Fundamental Plane of black hole activity by comparing results from time-averaged data against strictly simultaneous observations.}
   {Significant variability is detected in 86\% of the sample, showing a clear amplitude stratification across the spectrum. The fractional root mean square variability amplitude is highest in X-rays with a median of 30\% (11--67\%), followed by the optical $g$ and $r$ bands at 19\% (2--33\%) and 8.5\% (0.2--24\%), and the radio band at 10\% (4--23\%). The Mexican Hat analysis reveals a red-noise power spectrum where long-term fluctuations dominate. The sample follows the expected scaling of the Fundamental Plane where, despite individual sources shifting within the relation due to stochastic fluctuations, such variability-induced dispersion accounts for only $\sim$3\% of the total scatter. Our findings support a core-dominated origin for the 15 GHz emission, likely arising from a compact jet base or a magnetized corona, while reflecting intrinsic physical diversity across these AGN cores, where differences in variability patterns, radio-loudness, and Fundamental Plane location point toward distinct accretion/ejection processes and degrees of corona-jet coupling.}


\maketitle
\nolinenumbers

\section{Introduction}

Active Galactic Nuclei (AGN) are powered by the accretion of matter onto Super Massive Black Holes (SMBHs), generating intense emission that spans the entire electromagnetic spectrum, from radio to $\gamma$-rays \citep{netzer2013}. In the standard paradigm, the optical emission from accreting black hole systems is generally attributed to thermal radiation from an optically thick, geometrically thin accretion disk \citep{shakura1973}, while the X-ray emission arises from inverse Compton scattering of disk photons by a hot (10\textsuperscript{8–9} K), optically thin plasma, commonly referred to as the corona \citep{haardt1991}.

An intriguing aspect of AGN phenomenology is the wide range of their radio power. Historically, the AGN population has been divided into two primary categories based on their radio emission: Radio-Loud (RL) and Radio-Quiet (RQ) sources \citep{Kellermann1989}. In RL objects, the radio emission is dominated by powerful relativistic jets. However, RQ AGN constitute approximately 90\% of the total population, and the origin of their radio emission remains under debate. Proposed mechanisms include nuclear star formation, AGN-driven winds, small-scale or "failed" jets, or the corona itself \citep[see e.g.,][]{wilson1995, laor2008, zakamska2014, panessa2019}. 

Significant correlations between radio and X-ray emission suggest a deep coupling between inflowing accretion and outflowing ejection, leading to theories such as that the X-ray corona may be the jet base \citep{Markoff2005}.
Among them is the discovery of a relation linking X-ray luminosity, radio luminosity, and black hole mass, known as the Fundamental Plane (FP) of black hole activity, which extends across a wide range of accreting systems, from X-ray binaries to AGN \citep{merloni2003, falcke2004}.

The robustness of the FP has been further explored across diverse populations and cosmic scales. Recently, \cite{bariuan2022} examined a large archival sample of 353 quasars up to z$\sim$5, revealing a significant dichotomy in the parameters of the plane between RL and RQ sources. Their results suggest that while RL quasars are dominated by jet and disk emission, the RQ population remains more consistent with advection-dominated accretion flows. Establishing these relations often relies on non-simultaneous archival data, which can introduce artificial variance into the plane. While specific campaigns like the Fundamental Reference AGN Monitoring Experiment \cite[FRAMEx,][]{fischer2021} have begun utilizing simultaneous observations to refine the FP, the impact of variability on the observed scatter remains largely unexplored.

The importance of investigating radio variability in RQ AGN has only recently been recognized, as these fluctuations are typically modest compared to those in their RL counterparts \citep{anderson2005, mundell2009}. Nevertheless, synchronous monitoring across X-ray, optical, and radio bands has begun to reveal a complex, interconnected architecture. For instance, the coordinated variability between millimeter and X-ray emission in sources such as NGC~7469 \citep{behar2020} and MCG~+08-11-11 \citep{petrucci2023} supports a shared origin within the corona. In contrast, other systems show evidence of a disk-jet coupling through time lags, such as the 20-day delay observed in NGC~7213 \citep{bell2011}, or even inverse relationships where the radio flux dims as X-ray luminosity increases \citep{fernandez2022}.

This variability also provides direct evidence for the compact nature of these regions. In sources like Mrk~110, significant radio fluctuations on timescales of days to weeks constrain the emission to a region smaller than $180$ Schwarzschild radii, ruling out star formation or large-scale disk winds in favor of a low-power jet or an outflowing corona \citep{panessa2022b}. Building on this, \citet{chen2022} monitored a small sample of three RQ Seyferts (Mrk~110, Mrk~766, and NGC~4593) using quasi-simultaneous X-ray and radio data. Their detection of core variability with fractional amplitudes up to 9.5\%, combined with radio fluctuations lagging X-rays by several tens of days, further suggests that radio emission is physically tied to primary X-ray atmospheric heating or particle acceleration events.

These individual cases provide a vital precedent for the accretion-ejection link, and most studies to date have focused on a limited number of sources. While the connection between the accretion disk and the corona has been extensively documented with synchronous optical and X-ray variability, often interpreted through 'reprocessing' or 'intrinsic' fluctuations in the accretion flow \citep[e.g.][]{Lyubarskii1997, uttley2003, lira2015, Cackett2021, kara2025, Paolillo2025}, the relationship between these components and the compact radio emission remains poorly constrained. Consequently, there remains a significant need for multi-wavelength monitoring of larger samples over extended timescales to determine how these accretion-ejection signatures evolve across the broader AGN population.

In this work, we address these open questions by analyzing the multi-wavelength variability of a volume-limited sample of 14 local AGN. These sources were selected from the \textit{INTEGRAL}/IBIS catalog, ensuring a hard X-ray selected sample that minimizes biases related to obscuration. By combining radio monitoring from the Arcminute Microkelvin Imager (AMI-LA) with optical light curves from the Zwicky Transient Facility (ZTF) and X-ray data from \textit{Swift}/XRT, we characterize the variability properties across a broad range of the electromagnetic spectrum, from radio to X-rays.

The paper is organized as follows. In Section~\ref{sec:sample}, we describe the sample selection and the physical properties of the 14 AGN. Section~\ref{sec:observations} details the data reduction procedures for the radio, optical, and X-ray observations. Our variability analysis methodology is presented in Section~\ref{sec:analysis}, while the multi-wavelength results, including general variability features, radio-loudness parameters, and the FP, are presented in Section~\ref{sec:results}. In Section~\ref{sec:discussion}, we discuss our findings within the context of current literature. Finally, we summarize our main conclusions in Section~\ref{sec:conclusions}.


\begin{table*}
\centering
\small
\caption{\label{table:sample} Sample properties.}
\begin{tabular}{lcccccccccc}
\hline
Name & RA & DEC & Type & z & $M_{\rm BH}$ & $\lambda_{\rm Edd}$ & $L_X$ & $m_r$ & $L_R$ \\
 &  &  &  &  & (log $M_\odot$) & (log) & (log erg s$^{-1}$) & (mag) & (log erg s$^{-1}$) \\
\hline
IGR J00333+6122 & 00 33 18.41 & +61 27 43.1 & Sy1.5 & 0.105 & 8.54$^a$ & -1.89 & $43.15 \pm 0.27$ & $20.01 \pm 0.20$ & $39.83 \pm 0.06$ \\
QSO B0241+62 & 02 44 57.69 & +62 28 06.5 & Sy1 & 0.044 & 8.09$^a$ & -0.23 & $44.14 \pm 0.12$ & $15.76 \pm 0.09$ & $41.70 \pm 0.10$ \\
LEDA 168563 & 04 52 04.85 & +49 32 43.7 & Sy1 & 0.029 & 8.00$^a$ & -1.30 & $43.19 \pm 0.25$ & $16.60 \pm 0.13$ & $39.08 \pm 0.05$ \\
4U 0517+17 & 05 10 45.51 & +16 29 55.8 & Sy1.5 & 0.0179 & 7.00$^a$ & -0.24 & $43.24 \pm 0.19$ & $15.43 \pm 0.26$ & $38.43 \pm 0.03$ \\
MCG+08-11-11 & 05 54 53.61 & +46 26 21.6 & Sy1.5 & 0.0205 & 7.45$^a$ & -0.22 & $43.62 \pm 0.11$ & $14.16 \pm 0.29$ & $39.61 \pm 0.03$ \\
Mkn 3 & 06 15 36.36 & +71 02 15.1 & Sy2 & 0.0135 & 8.70$^a$ & -3.30 & $42.15 \pm 0.06$ & $13.85 \pm 0.01$ & $39.77 \pm 0.02$ \\
Mkn 6 & 06 52 12.25 & +74 25 37.5 & Sy1.5 & 0.0188 & 8.13$^a$ & -1.90 & $42.81 \pm 0.29$ & $14.15 \pm 0.24$ & $39.59 \pm 0.02$ \\
NGC 4388 & 12 25 46.75 & +12 39 43.5 & Sy2 & 0.0084 & 6.86$^a$ & -0.14 & $43.21 \pm 0.10$ & $15.50 \pm 0.01$ & $38.67 \pm 0.04$ \\
NGC 5252 & 13 38 16.00 & +04 32 32.5 & Sy1.9 & 0.023 & 9.00$^a$ & -1.96 & $43.47 \pm 0.09$ & -- & $39.12 \pm 0.10$ \\
2E 1853.7+1534 & 18 56 01.28 & +15 38 05.7 & Sy1 & 0.084 & 8.15$^b$ & -0.88 & $43.66 \pm 0.20$ & $19.25 \pm 0.08$ & $39.41 \pm 0.08$ \\
IGR J21247+5058 & 21 24 39.33 & +50 58 24.4 & Sy1 & 0.020 & 6.58$^c$ & 0.76 & $43.71 \pm 0.14$ & $15.69 \pm 0.02$ & $40.55 \pm 0.06$ \\
SWIFT J2127.4+5654 & 21 27 45.58 & +56 56 35.6 & NLS1 & 0.014 & 7.18$^d$ & -0.69 & $43.02 \pm 0.11$ & $20.16 \pm 0.11$ & $38.00 \pm 0.02$ \\
RX J2135.9+4728 & 21 35 54.02 & +47 28 22.3 & Sy1 & 0.025 & 7.35$^c$ & -0.78 & $43.09 \pm 0.15$ & $16.79 \pm 0.07$ & $38.51 \pm 0.04$ \\
IGR J23308+7120 & 23 30 37.28 & +71 22 46.0 & Sy2 & 0.037 & 7.68$^e$ & -2.16 & $42.24 \pm 0.13$ & $16.93 \pm 0.02$ & $38.29 \pm 0.08$ \\
\hline
\end{tabular}
\textit{Notes.}---Column (1): Source name; (2) and (3): Right ascension and declination; (4): Optical classification; (5): Redshift; (6): Black hole mass compiled from the literature: $^a$ \cite{chiaraluce2020}, $^b$ \cite{masetti2006}, $^c$ \cite{winter2010}, $^d$ \cite{malizia2008}, and $^e$ \cite{smith2020}; (7): Eddington ratio $\lambda_{\rm Edd}$ estimated in this work; (8): Mean unabsorbed X-ray luminosity in the 2--10\,keV band; (9): Mean optical magnitude in the $r$-band; (10): Mean radio luminosity at 15\,GHz. The values for $L_X$, $m_r$, and $L_R$ correspond to the mean values derived from the light curves presented in this work. $L_X$ was used to homogeneously estimate $\lambda_{\rm Edd}$ for all sources.
\end{table*}

\begin{figure*}
    \centering
    \includegraphics[width=\textwidth]{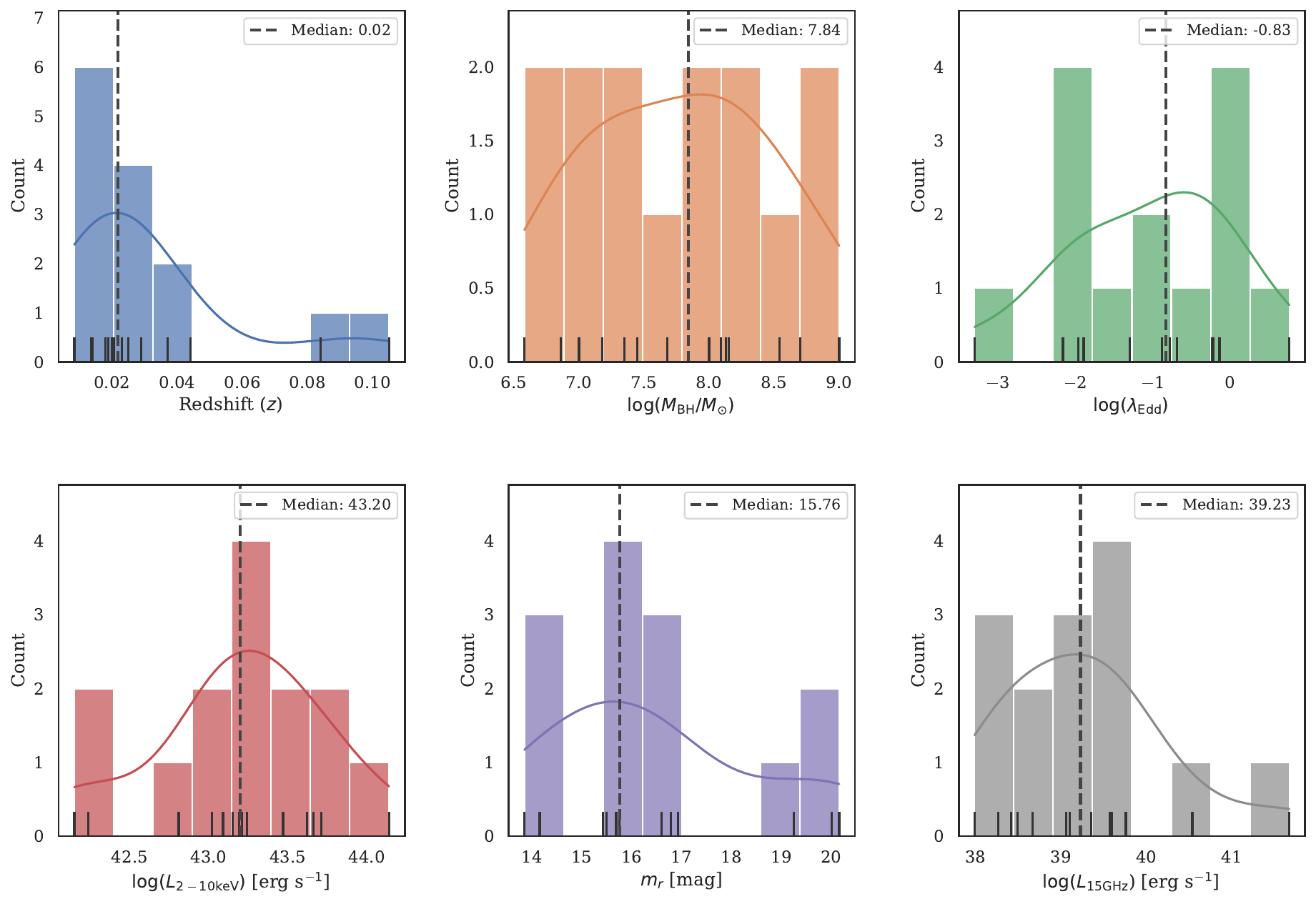}
    \caption{Distribution of the main physical properties for the 14 AGN in our sample. The panels show: (a) redshift ($z$), (b) black hole mass ($\log M_{\rm BH}$), (c) Eddington ratio ($\log \lambda_{\rm Edd}$), (d) hard X-ray luminosity ($\log L_{\rm 2-10 keV}$) from \emph{Swift}/XRT, (e) optical $r$-band magnitude ($m_r$) from ZTF, and (f) 15 GHz radio luminosity ($L_{\rm R}$) from AMI. In each panel, the solid line represents the Kernel Density Estimation (KDE), and the vertical dashed line indicates the median value of the distribution. Individual sources are marked with black ticks (rug plot) along the x-axis to illustrate the sample coverage.}
    \label{fig:sample_properties}
\end{figure*}

\section{Sample selection}
\label{sec:sample}


The starting point of our work is the sample of hard-X-rays selected objects presented in \cite{malizia2009}. It is a volume limited sample constituted by 88 nearby AGN extracted from a sample of 140 extragalactic objects from the INTEGRAL/IBIS \citep{bird2007}. This sample has several advantages: it is statistically complete and, being hard X-ray selected, it is relatively free of selection biases affecting samples selected in other ways, like UV-excess and IR (see discussion in \citealt{houlvestad2001}). It also has a wide coverage of information at X-rays and at multi-frequencies \citep{malizia2009, panessa2015, chiaraluce2020, panessa2022}. More importantly, it represents mainly high luminosity objects ($41.5\,<\,L_{2-10\,keV}\,<\,44.5$) with a wide range of X-ray radio-loudness parameter values (see \citealt{Terashima2003,chiaraluce2020}). The number of sources at DEC $\ge$ 0 deg, therefore observable with AMI, is 26, in which we already excluded BL LAC, Blazars and QSOs (following the optical classification of \citealt{masetti2012}). For our observing campaign, we calculated the expected 15 GHz flux density for each source by means of available data, mainly from the NRAO VLA Archive Survey (NVAS)\footnote{https://data.nrao.edu/}, and extrapolating from data at lower frequencies when no 15 GHz data were available\footnote{Assuming a power-law spectrum with a nominal spectral index of 0.5}. We excluded from the final sample the sources which have already been radio-monitored in the past, and those sources for which we could only provide upper limits on the 15 GHz flux density.
The final sample is constituted by 14 sources, and it is indicated in Table~\ref{table:sample}. The source names, coordinates (RA, Dec), redshifts ($z$), and black hole masses ($\log M_{\rm BH}$) were compiled from the literature (see references in Table~\ref{table:sample}). Notes on individual sources are reported in Appendix \ref{appendix:individualsources}.

The Eddington ratios, defined as $\lambda_{\rm Edd} = L_{\rm bol} / L_{\rm Edd}$, were self-consistently calculated for this study. We estimated the bolometric luminosity ($L_{\rm bol}$) from the hard X-ray luminosity ($2-10$\,keV) by applying a bolometric correction factor $\kappa_{2-10 \rm\, keV} \approx 20$, following the typical values for AGN at these scales \citep[e.g.,][]{Vasudevan2007, Lusso2012}. While more complex bolometric corrections exist \citep[e.g.,][]{duras2020}, we adopted a constant factor to ensure a homogeneous treatment across the entire sample, minimizing systematic offsets between individual sources. The Eddington luminosity was determined as $L_{\rm Edd} = 1.26 \times 10^{38} (M_{\rm BH}/M_\odot)$\,erg\,s$^{-1}$. 

The values for the hard X-ray luminosities ($L_{\rm X}$), optical $r$-band magnitudes ($m_r$), and radio flux densities ($L_{\rm R}$) correspond to the mean values derived from the light curves obtained in this work. A detailed description of the data reduction and the multi-wavelength variability analysis is provided in Section~\ref{sec:observations}.

The general physical properties and the parameter space covered by our sample are illustrated in Fig.~\ref{fig:sample_properties}. The distribution shows that our sample consists primarily of local AGNs ($z < 0.1$, median $z = 0.02$). The BH masses are well-distributed across nearly three orders of magnitude ($6.5 \lesssim \log M_{\rm BH} \lesssim 9.0$). The Eddington ratios show a broad range, from sub-Eddington sources ($\log \lambda_{\rm Edd} \approx -3.3$) to more active regimes. 

Through the whole manuscript, we assumed a flat $\Lambda$CDM cosmology with $H_0 = 70$ km s$^{-1}$ Mpc$^{-1}$, $\Omega_m = 0.3$, and $\Omega_\Lambda = 0.7$.

\section{Observations and data reduction}
\label{sec:observations}

The data used in this work includes dates between MJD [58400-59200], i.e., from October 2018 to December 2020, taken contemporaneously in the radio, optical and X-ray regimes. 
The typical cadence is approximately one observation per month in radio and X-rays, and one observation every three to four days in the optical.


\subsection{Radio AMI data}

The Arcminute Microkelvin Imager Large Array (AMI-LA) is an eight elements radio interferometer based in Cambridge, UK \citep{AMILA,AMILA2}. All eight antennas are 13m in diameter with a maximum baseline length of 110 m. It operates at a central frequency of 15.5\,GHz, with a 5\,GHz bandwidth allowing angular resolutions of 30 arcsec to be obtained. 

Observations were performed approximately monthly over the course of 2018$-$2020. Each observation was one hour long and the target was observed in 600~s scans, interleaved with 100~s observations of a nearby, bright phase calibrator. Daily observations of 3C~286 provide the necessary data for flux and bandpass calibration. To calibrate the data, we used a custom reduction package \textsc{REDUCE$\_$DC} \citep{Perrott2013,Perrott2015}, which flags for radio frequency interference (RFI) and performs standard radio calibration procedures. Imaging of the data was performed using the Common Astronomical Software Applications \textsc{CASA} software \citep{CASA} version 4.7. We removed any datasets where the images were poor due to antenna drop outs or large calibration errors. 

We used the \textsc{CASA} task \texttt{clean} to image the data with natural weighting, ascertained from the task \texttt{imstat}, including statistical uncertainties. We also imaged the phase calibrator and fit the source in order to estimate systematic variation in the flux bootstrapping process. We de-trended the phase calibrator fluxes with a second order polynomial to remove the year long trend and then estimate the variability in each flux calibrator. This variability ranged between 3$-$10$\%$, and was folded into the statistical target flux densities in quadrature.

\subsection{X-ray \emph{Swift}/XRT data}

We used data from a monitoring campaign starting in February 2019 and obtaining one snapshot observation of $\sim$1 ksec once per month up to November 2020 through ToO observations (PI: Chiaraluce). 

The data reduction of the {\it Swift} X-ray Telescope \citep[XRT,][]{2005SSRv..120..165B} in the Photon Counting mode was performed by following standard routines described by the UK Swift Science Data Centre (UKSSDC) and using the software in HEASoft version 6.28. Calibrated event files were produced using the routine \textsc{xrtpipeline}, accounting for bad pixels and effects of vignetting, and exposure maps were also created. Source spectra were extracted from circular regions with 30 arcsec and background spectra from free-of-sources regions with 80 arcsec radius located close to the nucleus. The \textsc{xrtmkarf} task was used to create the corresponding ancillary response files. The response matrix files were obtained from the HEASARC CALibration DataBase. The spectra were grouped to have a minimum of 20 counts per bin using the \textsc{grppha} task.

The X-ray data analysis was performed using XSPEC v. 12.11.1. 
All the spectra were fitted simultaneously to search for X-ray spectral variability \citep{lore2013}. We used an absorbed power law model (phabs*zwabs*powerlaw in XSPEC) and Galactic absorption for each of the sources was obtained from FTOOLS following \cite{1990ARAA..28..215D}. 
We notice that Mkn\,6 and 2E\,1853.7+1534 have a few observations with zero count-rate, so these were excluded from the analysis.

We let the column density, $N_H$, slope of the power law ($\Gamma$) and normalization of the power law vary for the different spectra of the same source. All of the sources show a better fit when the normalization varies. The $\chi^2$/d.o.f (degrees of freedom) are in the range [0.8-1.3]. The intrinsic fluxes were estimated from these models in two energy bands, the soft (0.2-2.0 keV), and the hard (2-10 keV).

Three of the sources (2E1853.7+1534, IGRJ23308+7120, and IGRJ00333+6122) did not have enough counts in the spectra to perform a spectral analysis. In these cases we used WebPIMMS\footnote{https://heasarc.gsfc.nasa.gov/cgi-bin/Tools/w3pimms/w3pimms.pl} to convert the count-rate from \emph{Swift}/XRT/PC to flux in the 0.2-10 keV energy band, using a power-law model with $\Gamma$=1.8, at the redshift and Galactic absorption for each source.

\subsection{Optical Zwicky Transient Facility (ZTF) data}
\label{sec:ztf}

We used $g$ and $r$-band light curves from the Zwicky Transient Facility \citep[ZTF;][]{graham2019} forced-photometry service \citep{masci2019, masci2023}. Source coordinates were obtained by cross-matching our sample with \textit{Gaia} EDR3 \citep{gaia2021} using a 1.5'' tolerance and confirmed through visual inspection of ZTF science, reference, and difference images. We note that NGC~5252 was excluded due to nuclear saturation. 
Additionally, the g-band data for IGR~J00333+6122, 2E~1853.7+1534, and SWIFT~J2127.4+5654 fall below the limiting magnitude and were therefore not used in the analysis. IGR~J23308+7120 was also excluded from the g-band analysis due to insufficient ZTF coverage (only 6 epochs available).
To ensure consistency with the multi-wavelength monitoring, data were restricted to the range MJD 58400--59200.

The forced-photometry data were cleaned following the procedures described in \cite{lore2023}. We applied several quality metrics, including \textsc{infobits}=0 and a limiting magnitude threshold ($r < 20.5$). We also implemented filter-dependent bad-data quality flags based on the photometric zero point (\textsc{zpmaginpsci}) and its RMS deviation (\textsc{zpmaginpscirms}) as detailed in the ZTF Public Data Releases. Specifically, we rejected epochs where $\textsc{zpmaginpsci} < \text{ZP}_{thres}[\text{rcid}] - C \cdot \text{sec}z$ (with $C=0.2$ for $g$ and $0.15$ for $r$) or $\textsc{zpmaginpscirms} > 0.06$ ($g$) and $0.05$ ($r$). Following these corrections, magnitudes were color-corrected using Pan-STARRS1 colors, and uncertainties were rescaled.

The observed magnitudes were corrected for Galactic extinction \citep{Schlafly2011}.
Since the forced photometry is performed on difference images, the host galaxy contribution is removed in the subtraction process, and the reference epoch flux is subsequently re-added to recover the total nuclear flux. No additional host-galaxy decomposition was applied beyond this. Furthermore, the data analysis was done through PSF photometry, ensuring that the flux comes primarily from the nuclear region in these very local galaxies. 

Since our primary focus is variability, we performed an additional quality control by comparing uncertainties with ZTF DR19. This led to the exclusion of Mkn~3 and $g$-band data for NGC~4388 and IGR~J21247+5058 (see Appendix~\ref{appendix:variability}). 

Finally, we cross-matched our sample with the \texttt{ALeRCE} broker \citep{forster2021} to identify sources with active detections in the ZTF alert stream. Six out of the 14 sources showed significant ($>5\sigma$) variations relative to their reference images: 
LEDA~168563 (ZTF18abvfsrv), 
MCG+08-11-11 (ZTF17aaaocpq), 
Mkn~6 (ZTF18abwbeui), 
RXJ2135.9+4728 (ZTF18abcnykp), 
4U0517+17 (ZTF18aaaejeh), and 
QSO~B0241+62 (ZTF18acvqufp).

\section{Variability analysis}
\label{sec:analysis}

\begin{figure*}[!ht]
    \centering
    \includegraphics[width=0.75\textwidth]{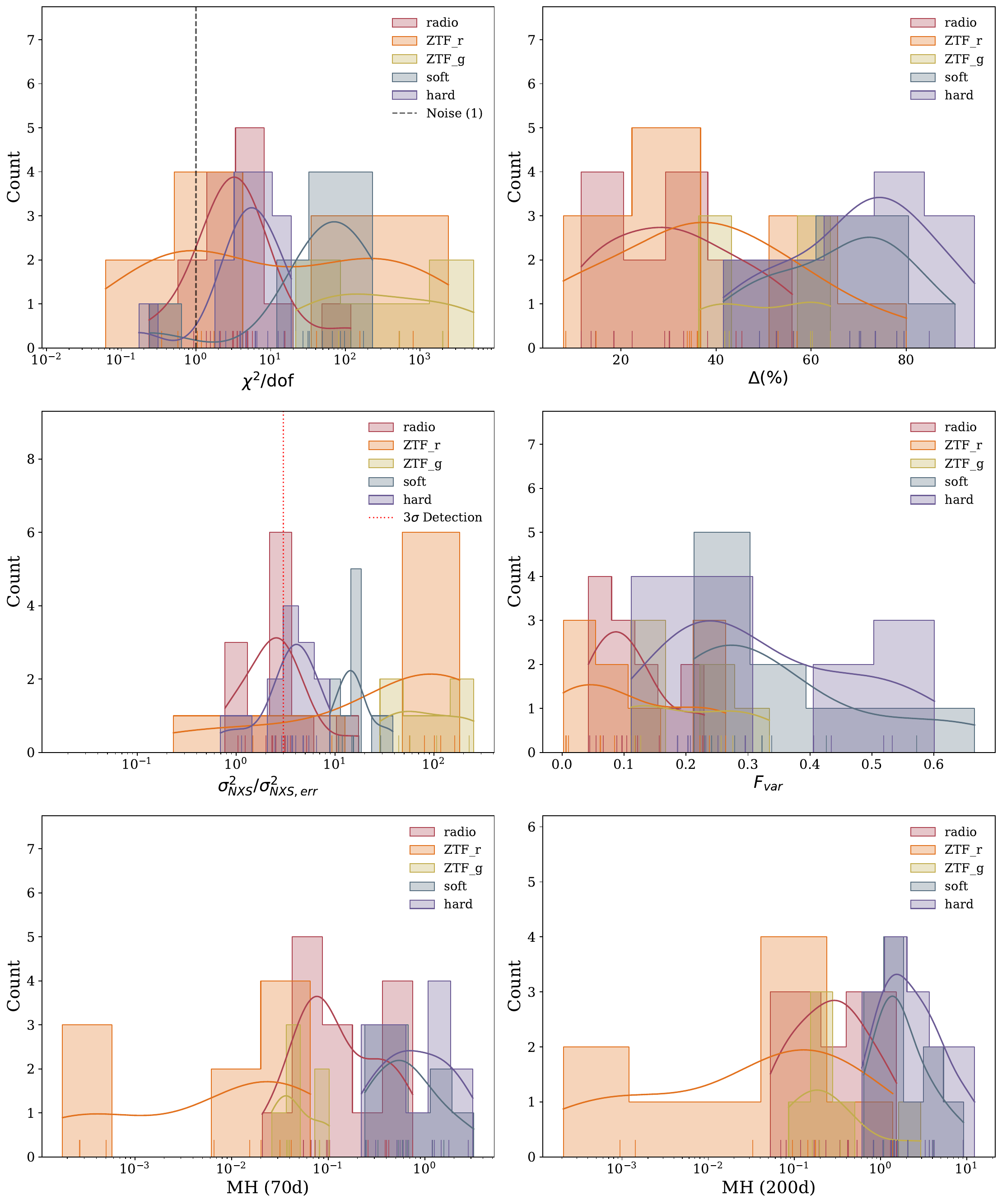}
    \caption{Distributions of variability features for the AGN sample across different energy bands (Radio, Optical $g$ and $r$, and Soft/Hard X-rays). 
    The panels show: (a) $\chi^2/\mathrm{dof}$, (b) relative change $\Delta (\%)$, (c) variability significance $\sigma^{2}_{NXS} / \sigma^{2}_{NXS,err}$, (d) fractional variability $F_{var}$, and (e-f) variance at 70 and 200 days time scales using the Mexican Hat (MH) filter. 
    Dashed black lines indicates the expected value for noise ($\chi^2/\nu = 1$), while the dotted red line in the significance panel marks the 3$\sigma$ detection threshold. Individual sources are indicated by rug plots at the base of each panel.}
    \label{fig:features_histograms}
\end{figure*}

The multi-band light curves for the sample are presented in Fig. \ref{fig:lcs_full}, where the individual flux variations for each source are displayed across the different energy regimes. The host galaxy was not subtracted from the optical data.

We show the results of the multi-band variability analysis in Table \ref{tab:variability_results} of the Appendix. 

For each of the observed bands, we report the $\chi^{2}$ of the light curve with respect to a constant flux model, along with the degrees of freedom (d.o.f.).
In addition, we have estimated the normalized excess variance, $\sigma^{2}_{NXS}$, and its associated error, $\sigma^{2}_{NXS,err}$, which represents the intrinsic variability amplitude of the light curves after accounting for measurement noise, following the prescriptions in \cite{vaughan2003}. 
It is worth noticing that the reported uncertainties account only for measurement noise. As discussed in \cite{vaughan2003} and \cite{allevato2013}, an additional scatter arises from the stochastic nature of the red-noise process, which depends on the intrinsic power spectral density (PSD) shape. Without repeated observations on the same timescales or dedicated simulations, this contribution cannot be formally quantified, and therefore the variability significance should be interpreted with caution. We note, however, that the lightcurves in different wavelengths for a given object are quasi-simultaneous, so if the variations were correlated they would be tracking the same realization of the variability process.
We also report the fractional root mean square (rms) variability amplitude, $F_{var}$, as defined in \cite{vaughan2003}, providing a normalized measure of the fluctuation scale.

Furthermore, we characterize the variability power at specific temporal scales using the Mexican Hat, following the methodology described in \cite{arevalo2012}. This technique is particularly well-suited for astronomical light curves as it provides a localized measure of the power spectrum in the time domain, offering significant advantages over traditional Fourier-based methods. Specifically, the Mexican Hat is robust against uneven sampling and is not affected by "red noise leakage", where power from scales longer than the total observation span distorts the high-frequency measurements. We calculated the variances at scales of 70 and 200 days, providing a direct estimate of the power concentrated at these frequencies and allowing for a comparison of short- versus long-term fluctuation amplitudes.
The timescales of 70 and 200 days were chosen to be well-sampled by our monthly cadence  observations over the two-year monitoring period. The shortest timescale we probe includes at least two data points per cycle and the longest is chosen to be contained 2-3 times within the length of the light curve. This provides a robust characterization of both short- and long-term variability. The variances are calculated from filtered lightcurves that have been normalized to the corresponding mean flux. Therefore the variances are relative to these mean values and can be compared across different wavebands and objects. 

The limitations in sampling cadence prevent a reliable determination of significant time delays between the radio, optical, and X-ray bands.

\section{Results}
\label{sec:results}

\subsection{Histograms of the features}

The multi-wavelength variability analysis, summarized in the distributions shown in Fig.~\ref{fig:features_histograms}, reveals a high degree of intrinsic activity across the sample. The statistical significance of these variations is confirmed by the $\chi^2/\mathrm{dof}$ and $\sigma^{2}_{NXS}/\sigma^{2}_{NXS, err}$ distributions, where the majority of sources---particularly in the X-ray and optical bands---lie well beyond the expected noise levels and the 3$\sigma$ detection threshold. This indicates that the observed flux changes are dominated by physical processes rather than instrumental fluctuations. 

\begin{table}[h]
\centering
\footnotesize 
\setlength{\tabcolsep}{3pt} 
\begin{tabular}{lccccccr}
\hline \hline
Source & radio & $g$ & $r$ & S & H & Var. & Det. \\ \hline
IGR\,J00333+6122 & 2.3 & -- & $<$0 & -- & 3.7 & N - N - Y & 1 \\
QSO\,B0241+62 & 5.1 & 57.4 & 100.9 & 17.5 & 4.0 & YYYYY & 5 \\
LEDA\,168563  & 3.2 & 44.0 & 80.3 & 10.8 & 3.3 & YYYYY & 5 \\
4U0517+17 & 1.2 & 106.2 & 117.4 & 11.9 & 4.7 & NYYYY & 4 \\
MCG+08-11-11 & 3.1 & 227.7 & 181.8 & 15.4 & 8.9 & YYYYY & 5 \\
Mkn\,3 & $<$0 & -- & -- & -- & $<$0 & N - - - N & 0 \\
Mkn\,6 & 2.3 & 252.4 & 162.0 & 38.4 & 5.5 & NYYYY & 4 \\
NGC\,4388 & 3.6 & -- & 0.2 & 2.0 & 2.9 & Y - NNN & 2 \\
NGC\,5252 & 17.3 & -- & -- & 15.2 & 0.7 & Y - - YN & 2 \\
2E\,1853.7+1534 & 2.5 & -- & $<$0 & -- & 6.5 & N - N - Y & 1 \\
IGR\,J21247+5058 & 2.0 & -- & 9.3 & 28.9 & 4.8 & N - YYY & 3 \\
SWIFT\,J2127.4+5654 & $<$0 & -- & $<$0 & 14.8 & 2.5 & N - NYN & 1 \\
RX\,J2135.9+4728 & 0.8 & 28.5 & 56.7 & 15.5 & 3.7 & NYYYY & 4 \\
IGR\,J23308+7120  & 1.0 & -- & 1.5 & -- & 1.1 & N - N - N & 0 \\ \hline
\end{tabular}
\caption{Significance of variability, $S$. Columns refer to: Radio, ZTF-$g$, ZTF-$r$, S and H refer to soft and hard X-rays. Var. means that variability is detected (Y) or ot (N) in each band, respectively, and Det. refers to the number of variable bands. Bands with $<$0 mean that only upper limits of $ \sigma_{NXS}^2$ was estimated. }
\label{tab:sig_summary}
\end{table}

The statistical significance of the intrinsic variability was evaluated using the ratio $S = \sigma^{2}_{NXS} / \sigma^{2}_{NXS, err}$ (Fig.~\ref{fig:features_histograms}, middle-left), adopting a $3\sigma$ threshold for formal detections. As summarized in Table~\ref{tab:sig_summary}, 12 out of 14 sources (86\%) exhibit significant variability in at least one energy band. The objects QSO\,B0241+62, LEDA\,168563, and MCG+08-11-11 are the most variable in our sample, showing significant detections across all five spectral regimes (Det.=5). 
In contrast, 4U\,0517+17, Mkn\,6, IGR\,J21247+5058 and RX\,J2135.9+4728 exhibit strong multi-band variability in X-rays and optical but variations in the radio cannot be confirmed (Det.=3-4). 
Variations only in the X-ray regime were detected in IGR\,J00333+6122, 2E\,1853.7+1534, and SWIFT\,J2127.4+5654 (Det.=1).
NGC 4388 is the only source where variations are found in the radio band alone.
Only IGR\,J23308+7120 and Mkn\,3 showed no significant variability across the entire monitored period, although we notice that Mkn\,3 had no useful optical data.
Finally, NGC\,5252 shows variations in the X-rays and radio, but there is no optical data.

The fractional variability ($F_{var}$) analysis reveals a clear stratification across the electromagnetic spectrum. As expected, the X-ray regime exhibits the most pronounced fluctuations, with values ranging from 11.2\% to 60.1\% in the hard band (median $\sim$29.5\%) and reaching a maximum of 66.6\% in the soft band (median $\sim$30.1\%). In the optical domain, we find moderate variations with medians of 8.5\% and 19.2\% for the ZTF $r$ and $g$ bands, respectively, with a maximum detected amplitude of 33.4\%. Finally, the radio band shows relatively more stable behavior, with a median of 10.0\% and a total range spanning from 4.3\% to 22.9\%.
The analysis of the Mexican Hat at different time scales is consistent with these results and shows also that the power increases from 70 to 200 days across all bands (see next Section for details).

We performed a correlation analysis between the variability estimators across different energy bands (e.g., $F_{var}$ in radio vs. optical) and investigated their potential dependence on fundamental physical properties, such as black hole mass ($M_{BH}$) and Eddington ratio ($\lambda_{\rm Edd}$). In all cases, no statistically significant correlations were found (p$>$0.05). This lack of correlation may be attributed to the relatively small size of our sample, which limits the statistical power to detect subtle trends. Additionally, the discrepancy in sampling cadences means that each band probes different frequency domains of the power spectral density, potentially masking any simple linear relationship between the variability amplitudes.

\subsection{Mexican Hat}

\begin{figure}
    \centering
    \includegraphics[width=0.45\textwidth]{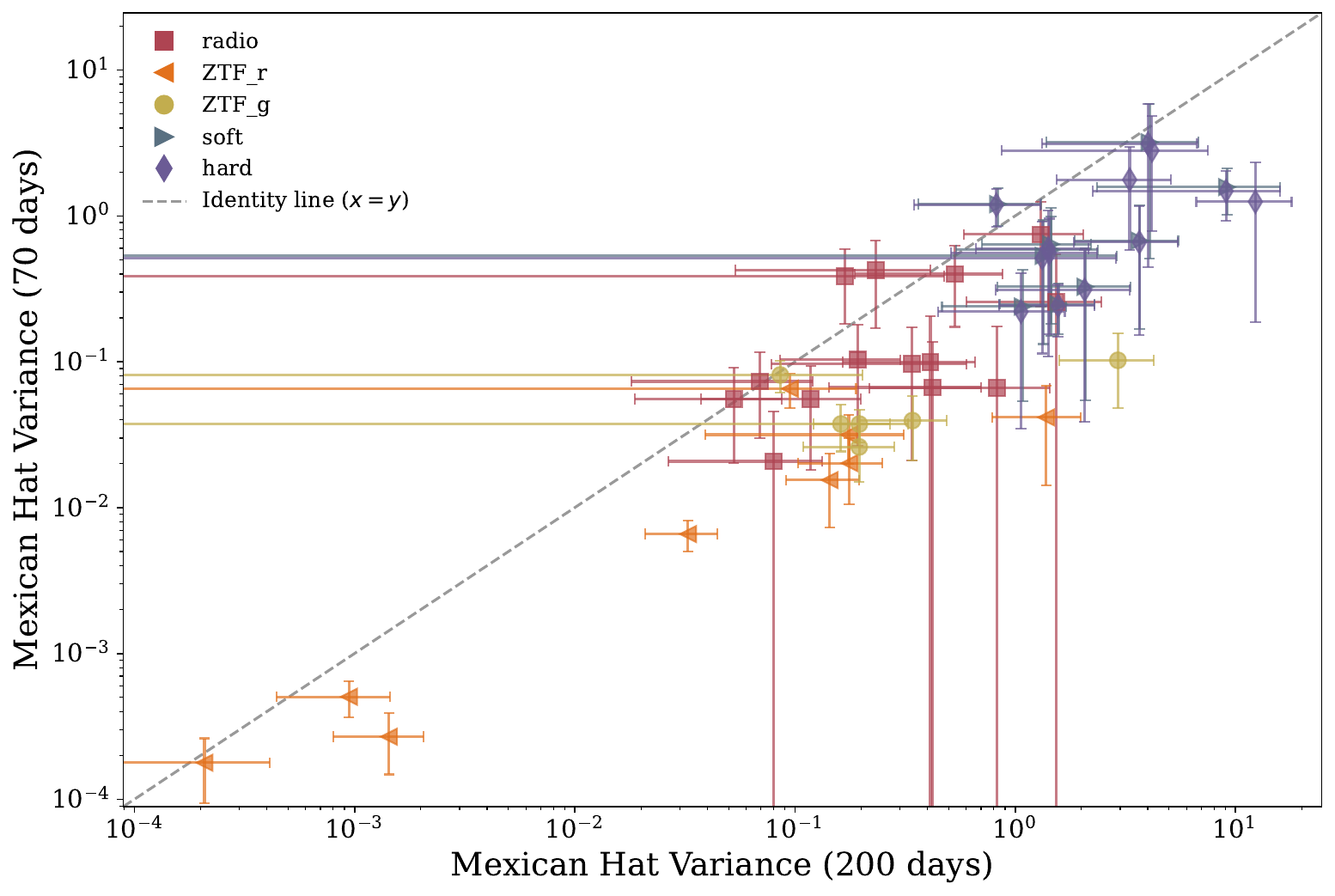}
    \caption{Comparison of the Mexican Hat wavelet variance at two different time scales: 200 days versus 70 days. The dashed black line represents the 1:1 identity relation. Sources lying below this line exhibit more power at longer time scales, consistent with the characteristic red-noise stochastic variability of AGNs. Data points are color- and marker-coded by energy band.}
    \label{fig:mexican_hat_plot}
\end{figure}

The distribution of the Mexican Hat variances at 70 and 200 days across the sampled AGN population is presented in Fig.~\ref{fig:mexican_hat_plot} and reveals a clear hierarchy of variability that is consistent with the stochastic nature of accretion processes. A primary observation is that the majority of sources lie below the 1:1 identity line, indicating that the power at the 200-day scale consistently exceeds that of the 70-day scale. This is a characteristic signature of ``red noise'' behavior, where long-term fluctuations carry significantly more power than short-term ones, following a power-law PSD shape, $P(f) \propto f^{-\alpha}$ \citep{vaughan2003}.

These data show a distinct energy-dependent stratification, where X-ray emissions (hard and soft bands) are systematically shifted toward the upper right of the diagram. This high-amplitude, high-frequency variability is a hallmark of the X-ray corona. Because the emission region is physically small, light-crossing times are short, allowing for the rapid, high-amplitude fluctuations observed here \citep{uttley2002, mchardy2006}. In contrast, the optical bands ($g$ and $r$) occupy a lower-variance space, consistent with the standard accretion disk model where the optical emission originates from more extended regions. Consequently, the disk acts as a ``low-pass filter'' where the geometric extent of the emitting region smooths out rapid fluctuations through averaging \citep{mushotzky2011}. 
The optical r-band spans the widest dynamic range of variability, covering nearly four orders of magnitude. This is primarily due to the inclusion of type 2 sources, which exhibit significantly lower variability and occupy the lower end of the relation. 
In contrast, the narrower distribution in the g-band reflects a selection effect, since the six sources with g-band data are all Seyfert 1.x, as the Seyfert 2 sources in the sample lack reliable g-band measurements (see Section \ref{sec:ztf}).

The radio band shows the highest scatter around the 1:1 identity line and the largest individual uncertainties. Unlike the optical bands, the radio data points do not follow a clear trend along the identity line, a behavior that can be attributed to the complex nature of radio emission in AGNs  (see Section \ref{sec:discussion}).

Furthermore, the large error bars suggest that many radio observations may be approaching instrumental sensitivity limits, introducing a noise floor that further obscures any intrinsic physical signal.

\subsection{Radio loudness}

We estimated the radio-loudness of the sample using two complementary approaches: one relative to the optical flux and another relative to the X-ray luminosity.
While the classical definition of $R_{\rm opt}$ by \citet{Kellermann1989} originally employed radio data at 5~GHz and optical $B$-band magnitudes, we adapt this diagnostic to our dataset by using the mean values of the 15~GHz radio fluxes and ZTF $r$-band photometry during this monitoring. This choice is motivated by the simultaneity of our multi-wavelength coverage, which minimizes uncertainties related to AGN variability. For a typical flat-spectrum radio core, characterized by a spectral index $\alpha \approx 0$ (where $F_\nu \propto \nu^\alpha$), the flux density at 15~GHz is a reliable proxy for the 5~GHz emission \citep{Ho2001, Nagar2005}. For sources with steep radio spectra ($\alpha > 0.5$), our 15~GHz-derived $R_{\rm opt}$ should be considered a conservative lower limit. At the traditional 5~GHz frequency, these sources would appear more RL.
In general, by adopting 15~GHz instead of the classical 5~GHz, we preferentially probe the compact, self-absorbed regions of the jet. This minimizes the contribution from extended, relic emission, providing a more contemporaneous view of the central engine's activity.
Furthermore, the $r$-band is significantly less affected by both Galactic and host-galaxy extinction compared to the $B$-band, as the extinction $A_\lambda$ decreases at longer wavelengths \citep{Cardelli1989, Schlafly2011}, providing a more robust estimate of the nuclear continuum in our sample. 
Nevertheless, r-band measurements may be affected by host-galaxy dilution, especially in type 2 sources where the nuclear emission is obscured, potentially resulting in lower limits for the calculated radio-loudness.
We note that $R_{\rm opt}$ could not be determined for NGC~5252 because its nuclear r-band emission is saturated.

We define the radio-loudness parameter as the ratio between the radio flux density and the optical flux density:

\[
R_{\rm opt} = \frac{F_{\rm 15\,GHz}}{F_r},
\]

\noindent where $F_{\rm 15\,GHz}$ is the radio flux at 15 GHz, converted from mJy to Jansky, and $F_r$ is the extinction-corrected flux density in the r-band (AB magnitude). 

It is worth noting that while the traditional thresholds for radio-loudness ($\log R_{\rm opt} > 1$ and $\log R_X > -2$) were originally calibrated using 5~GHz data \citep{Kellermann1989, Panessa2007}, we maintain these criteria as a robust benchmark for our 15~GHz study. Given that the majority of our sample exhibits flat or steep radio spectra ($\alpha \gtrsim 0$; see Table~\ref{tab:radio_loudness}), the use of a higher frequency acts as a conservative filter that targets the compact nuclear core while minimizing extended emission. Consequently, any source exceeding these classical limits at 15~GHz provides a particularly stringent and reliable indication of a powerful, active jet.

Following the literature, objects with $\log R_{\rm opt} < 1$ are considered RQ, while those with $\log R_{\rm opt} > 1$ are considered RL \citep{Kellermann1989,Panessa2007}.

We also calculated a radio-loudness parameter relative to the X-ray luminosity, as introduced by \citet{Terashima2003} and \citet{Panessa2007}:

\[
R_X = \frac{\nu L_\nu(\rm radio)}{L_X},
\]

\noindent where $L_X$ is the intrinsic 2--10 keV X-ray luminosity (in erg s$^{-1}$), $\nu = 15$ GHz, and $L_\nu({\rm radio})$ is the radio luminosity.

The specific thresholds used to distinguish between these populations have evolved as larger and more sensitive samples were analyzed. \citet{Terashima2003} originally proposed a criterion of $\log R_X > -4.5$ to identify RL sources. However, this limit was later refined by \citet{Panessa2007}, who introduced a more stringent threshold of $\log R_X > -2$. 
Under this revised classification, sources with $\log R_X < -4$ are strictly considered RQ, while we refer to those falling between these two limits as 'radio-intermediate', representing a transitional population with moderate radio-to-X-ray ratios.

\begin{table}
\centering
\tiny 
\setlength{\tabcolsep}{3pt} 
\caption{Radio-loudness parameters for the sample. Labels $R_{\rm opt}$ and $R_{\rm X}$ correspond to the optical and X-ray definitions, respectively. $R_{\rm opt}$P and $R_{\rm X}$P correspond to the values reported in \cite{panessa2022}, as well as their spectral index, $\alpha$, between the 5-10/10-15/15-22 GHz bands. }
\label{tab:radio_loudness}
\begin{tabular}{lcccccc}
\hline \hline
Source & Type & $\log R_{\rm opt}$ & $\log R_{\rm X}$ & $\log R_{\rm X}$P & $\log R_{\rm opt}$P & $\alpha$ \\ 
\hline
IGR J00333 & Sy1.5 & 0.53 & -3.32 & -4.42 & 1.90 & -/-/- \\ 
QSO B0241 & Sy1 & 1.95 & -2.44 & -3.06 & 0.70 & -0.86/-1.34/0.25 \\ 
LEDA 168563 & Sy1 & 0.03 & -4.11 & -4.70 & 1.50 & -/-/-  \\ 
4U 0517+17 & Sy1.5 & -0.27 & -4.81 & -5.27 & -  & -/-/- \\ 
MCG+08-11-11 & Sy1.5 & 0.36 & -4.01 & -4.25 & 1.27 & 1.07/1.1/0.14  \\ 
Mkn 3 & Sy2 & 0.79 & -2.39 & -4.79 & 1.03 &  1.41/0.35/1.3 \\ 
Mkn 6 & Sy1.5 & 0.48 & -3.23 & -3.93 & 1.17  & 0.91/1.0/- \\ 
NGC 4388 & Sy2 & 0.91 & -4.54 & -4.94 &  -0.80  & 0.9/0.8/1.08 \\ 
NGC 5252 & Sy1.9 & --- & -4.35 & -4.71 & 0.60  & 0.04/0.61/0.28 \\ 
2E 1853 & Sy1 & 0.27 & -4.26 & - &  -  & -/-/-  \\ 
IGR J21247 & Sy1 & -0.08 & -3.16 & -  & -  & -/-/- \\ 
Swift J2127 & NLS1 & 0.51 & -5.02 & -  & - & -/-/-  \\ 
RX J2135 & Sy1 & -0.25 & -4.59 & -  & - & -/-/-  \\ 
IGR J23308 & Sy2 &  -0.77 & -3.95 & -5.16 & 1.12  & 0.64/-0.4/- \\ 
\hline
\end{tabular}
\end{table}

It is important to stress that $R_{\rm opt}$ and $R_X$ measure radio-loudness in different regimes and are not directly comparable in absolute terms. While $R_{\rm opt}$ compares radio to optical emission and is more sensitive to the host galaxy and nuclear accretion disk emission, $R_X$ compares radio to X-ray emission, tracing the relative strength of the jet compared to high-energy corona emission.
Radio loudness parameters are presented in Table \ref{tab:radio_loudness}.
A comparison with \cite{panessa2022} reveals generally lower $R$ values in our study. This shift is likely a combination of: (i) the use of $r$-band photometry, which recovers nuclear emission obscured in the $B$-band, and (ii) the 15~GHz frequency targeting the most active, compact core. For the steep-spectrum sources in our sample (see $\alpha$ in Table~\ref{tab:radio_loudness}), our estimates represent a lower limit of the radio-loudness.

\begin{figure}
    \centering
    \includegraphics[width=0.5\textwidth]{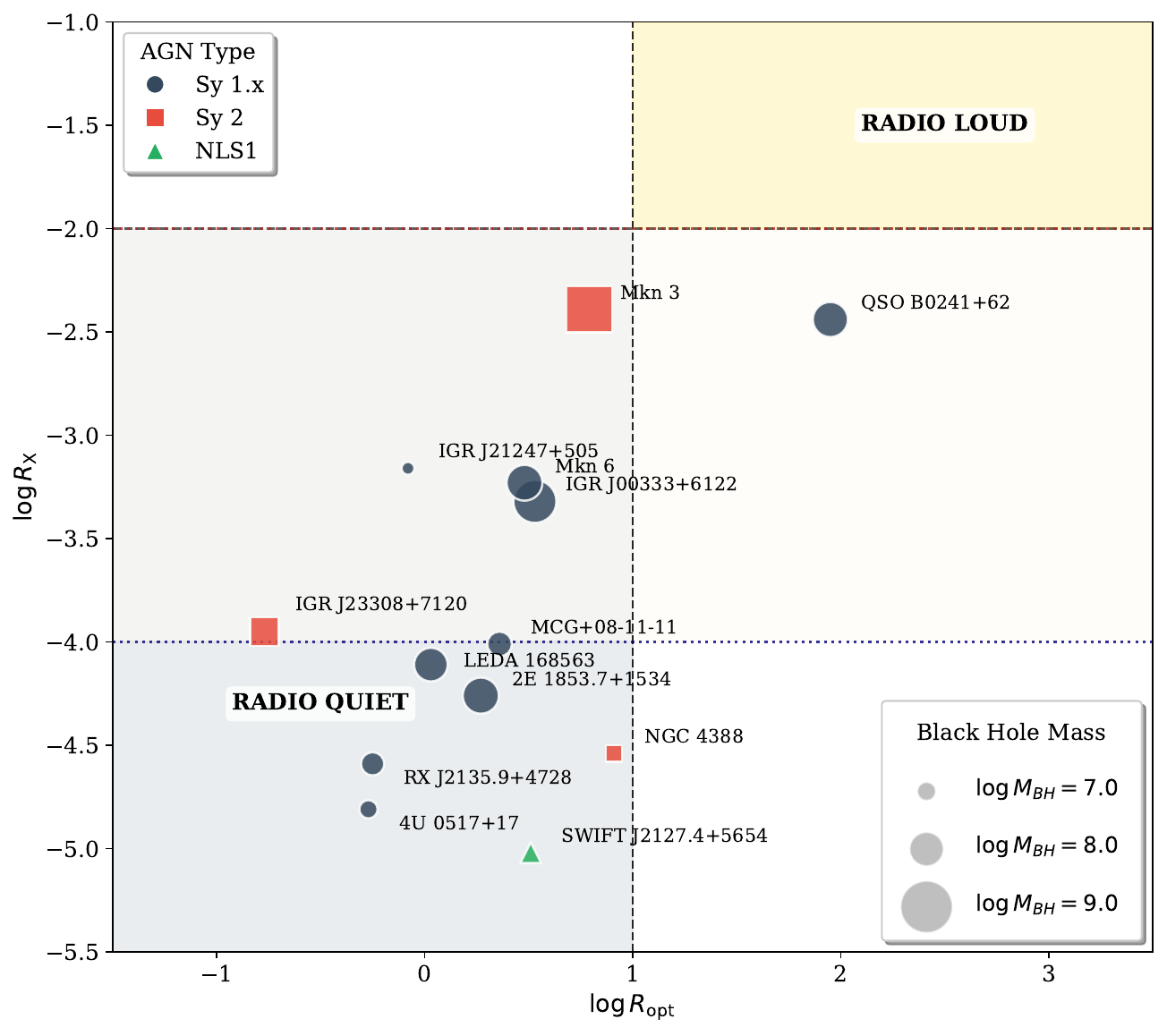}
    \caption{
    Comparison between optical, $R_{\rm opt}$, and X-ray, $R_{\rm X}$, radio-loudness parameters for the sources in the sample. 
    The vertical dashed line at $\log R_{\rm opt}=1$ marks the conventional boundary between RL and RQ sources in the optical definition, whereas the horizontal dashed line at $\log R_X=-2$ and $\log R_X=-4$ indicates the threshold commonly adopted to identify RL AGN in X-rays \citep[e.g.][]{Panessa2007}.
    Different colors denote the optical classification of the sources (Sy~1, Sy~2, and NLS1). The size of the symbols is proportional to $M_{BH}$.
    }
    \label{fig:radio_loudness_diagnostic}
\end{figure}

The fraction of RL sources in our sample varies depending on the chosen classification proxy. According to the optical radio-loudness parameter ($\log R_{\rm opt} > 1$), the RL fraction is 8\% (1/13). In contrast, when using the X-ray radio-loudness parameter ($\log R_X > -2$), the RL fraction drops to 0\% (0/14). Instead, 43\% of the sample (6/14) is classified as radio-intermediate ($-4 < \log R_X < -2$), while the remaining 57\% (8/14) are strictly radio-quiet ($\log R_X < -4$).

By examining the black hole masses ($M_{BH}$) in the $R_{\rm opt}$--$R_X$ plane (see Fig.~\ref{fig:radio_loudness_diagnostic}), we find that sources in the radio-intermediate regime tend to have slightly higher masses, with a median $\log (M_{BH}/M_{\odot}) = 8.11$ compared to $7.40$ for the strictly radio-quiet subsample. However, a Kolmogorov-Smirnov test indicates that this difference is not statistically significant ($p = 0.38$). While these results might be visually consistent with the 'two-sequence' model where radio-loudness scales with $M_{BH}$ \citep{Sikora2007}, a larger sample would be required to statistically confirm this trend.

\subsection{The Fundamental Plane of Black Hole Activity}

The Fundamental Plane (FP) of black hole activity represents a fundamental scaling relation between the disk-jet coupling and the mass of the central engine. Originally proposed by \cite{merloni2003} and \cite{falcke2004}, this non-linear correlation relates the radio luminosity ($L_R$), the X-ray luminosity ($L_X$), and the black hole mass ($M_{BH}$) through the expression:
\begin{equation}
    \log L_R = \xi \log L_X + \eta \log M_{BH} + b
\end{equation}
where $\xi$ and $\eta$ are the scaling coefficients for the X-ray luminosity and mass, respectively, and $b$ is a constant offset. This relation suggests that the physical processes governing accretion and jet formation are scale-invariant across the entire black hole mass spectrum, from stellar-mass black holes to supermassive black holes in AGN.

In this work we explore the FP using quasi-simultaneous X-ray and radio data to minimize the impact of intrinsic source variability, requiring radio and X-ray data to be within a temporal window of $\pm 1$ day. We used different measurements of the same source to also study the effect of variability in the dispersion of this plane.

The sample was initially compared to the classical \cite{merloni2003} relation using the standard projection $\log L_R - 0.78 \log M_{BH}$ vs. $\log L_X$. To further investigate the population dichotomy, we adopted the recent calibration by \cite{bariuan2022}. Following their methodology for Quasars, we projected our data onto a coordinate system defined by $1.44 \log L_X + 0.16 \log M_{BH}$ in the abscissa and $\log L_R$ in the ordinate. This specific scaling highlights the differences between RL and RQ populations.
We incorporated four distinct models from \cite{bariuan2022}: the local ($z < 1.5$) and global (all redshift) relations for both RL and RQ populations. For the local sample ($z < 1.5$), the coefficients used were $\xi = 1.32, \eta = -0.06, b = -16.20$ for RL sources, and $\xi = 0.38, \eta = 0.40, b = 20.20$ for RQ sources.

\begin{figure*}
\centering
\includegraphics[width=0.49\textwidth]{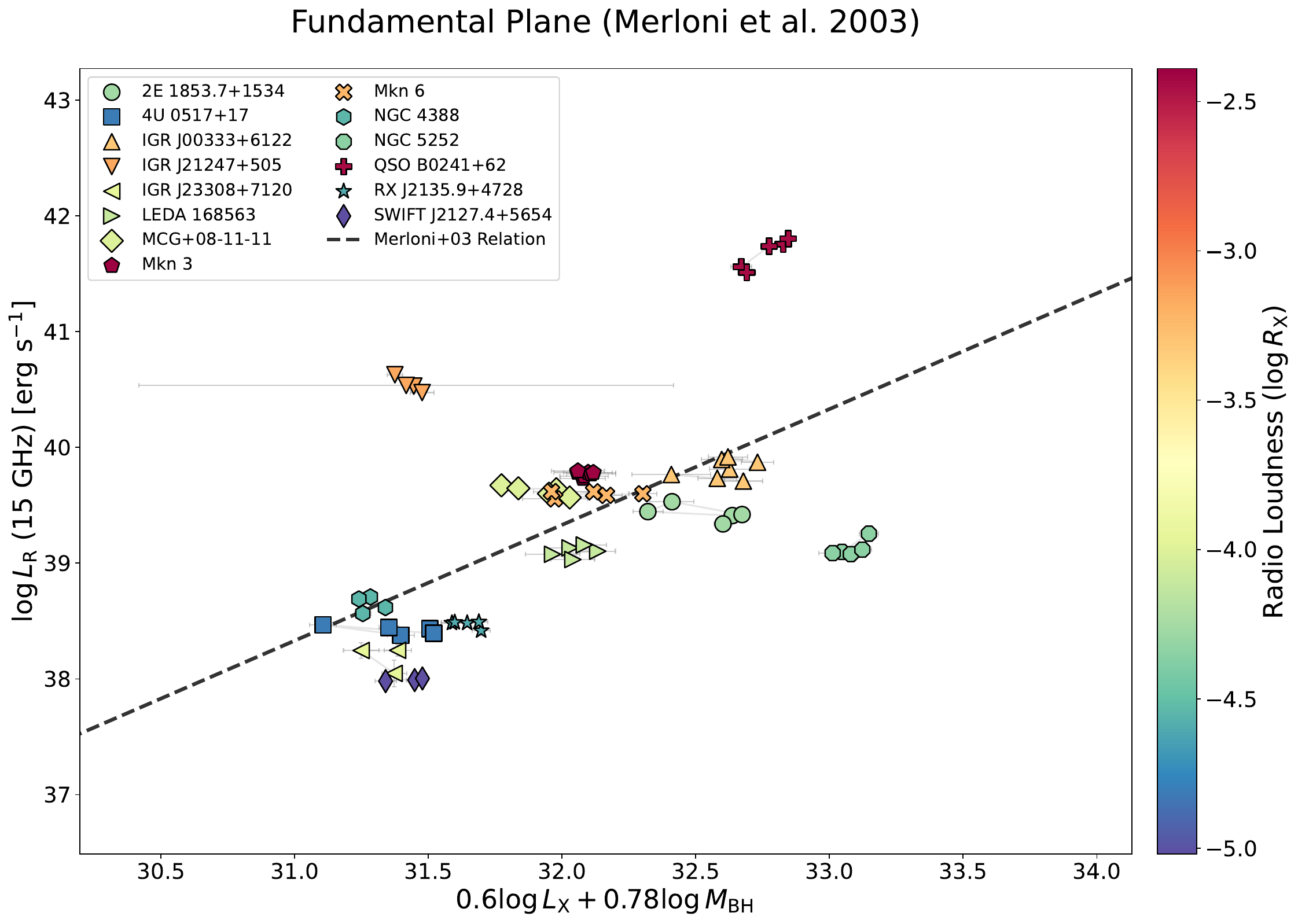}
\includegraphics[width=0.49\textwidth]{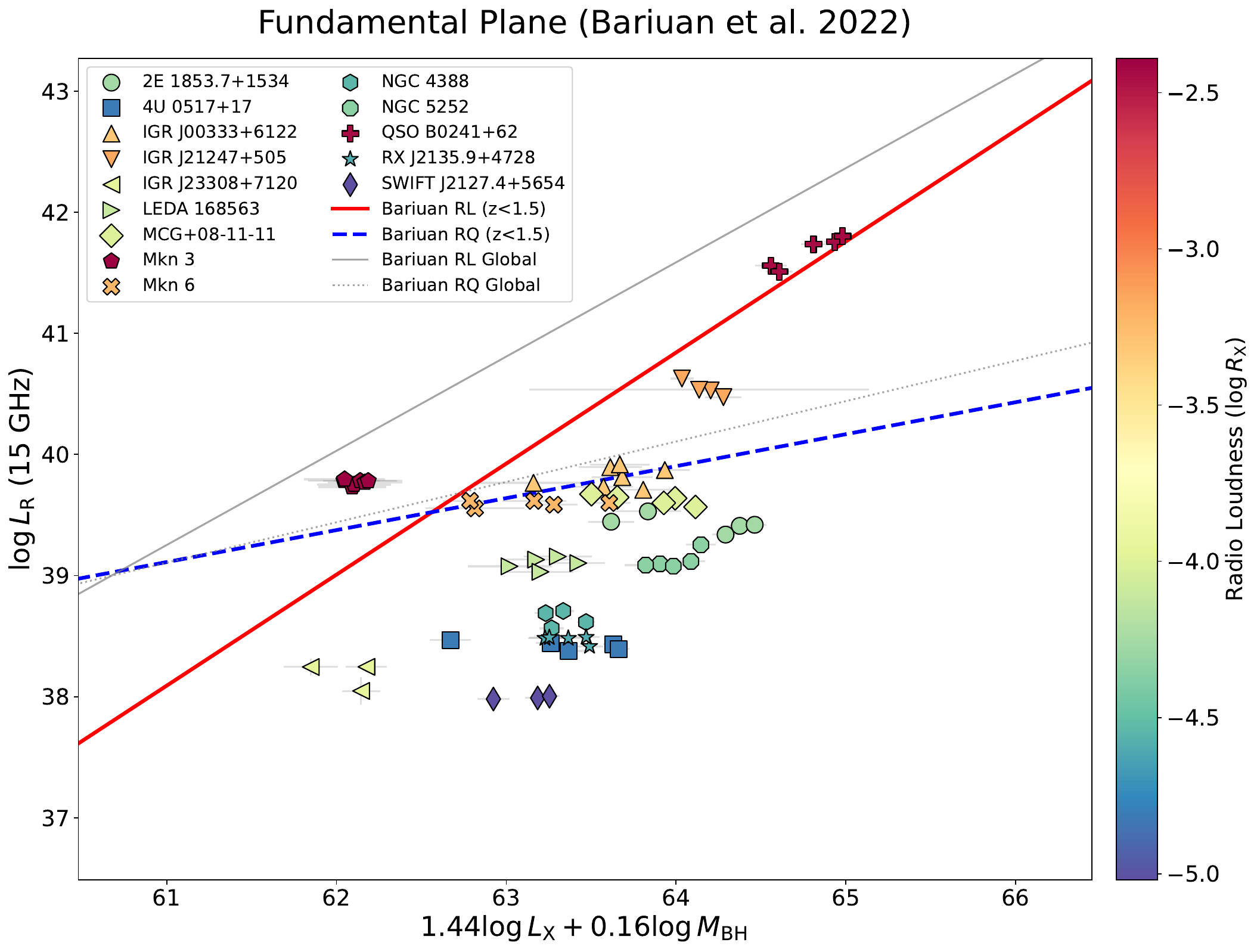}
\caption{Fundamental Plane of black hole activity for the AGN sample. Left: comparison with \cite{merloni2003} scaling. Right: comparison with \cite{bariuan2022}. Each source consists of multiple data points, each corresponding to a different epoch with simultaneous radio and X-ray observations. Points are color-coded by radio loudness $\log R_{\rm X}$.}
\label{fig:fp_comparison}
\end{figure*}

The results are presented in Fig. \ref{fig:fp_comparison}. To quantitatively evaluate the performance of each scaling relation, we defined the residuals as $\Delta \log L_R = \log L_{R,obs} - \log L_{R,pred}$. The scatter for the entire sample was calculated as the standard deviation of these residuals ($\sigma_{tot}$), while the systematic offset was quantified using the mean bias. Additionally, to isolate the impact of intrinsic source variability, we computed the internal variance for each source with multiple epochs ($N \geq 2$). 
We defined the variability-induced dispersion as $S_{var} = \sqrt{(\xi \sigma_X)^2 + \sigma_R^2}$, where $\xi = 0.6$ is the scaling coefficient from \cite{merloni2003}. The fractional contribution of variability to the total observed scatter was then determined by the ratio of the mean internal variance to the global variance ($\sigma^2_{var} / \sigma^2_{tot}$).

Our results indicate that the \cite{merloni2003} relation provides the most accurate description of our sample, yielding a near-zero mean bias of $0.027$~dex and a total scatter of $0.786$~dex. In contrast, the \cite{bariuan2022} relations exhibit significant systematic offsets, with biases of $-0.35$~dex and $-0.93$~dex for the RQ and RL models, respectively.

A detailed analysis of the individual bands reveals that the X-ray component (scaled by $\xi$) accounts for a mean contribution of $1.44\%$ to the total sample variance, while the radio component represents $0.73\%$. Combined, the total variability-induced dispersion ($\sigma_{var} = 0.118$~dex) accounts for only $2.16\%$ of the global observed scatter. At the individual level, the highest contributions are observed in \object{4U~0517+17} ($4.72\%$) and \object{2E~1853.7+1534} ($4.61\%$), primarily driven by X-ray fluctuations. The maximum radio-driven contribution is found in \object{QSO~B0241+62} ($2.72\%$).

Regarding individual residuals, we identify significant outliers such as \object{IGR~J21247+505} ($+1.78$~dex), \object{QSO~B0241+62} ($+1.58$~dex), and \object{NGC~5252} ($-1.29$~dex). Excluding these cases, the majority of the sample remains tightly clustered, with residuals typically below $0.5$~dex. The negligible mean bias and the fact that variability-induced scatter accounts for less than $3\%$ of the total variance suggest that the observed dispersion is not a result of observational uncertainty or source stochasticity. Instead, the residuals likely represent intrinsic physical departures from the scaling relation, potentially reflecting differences in accretion efficiency or jet properties across the sample.

\section{Discussion}
\label{sec:discussion}

In this work, we present multi-wavelength monitoring for a sample of 14 hard X-ray selected AGNs. By combining 15 GHz radio data with quasi-simultaneous optical and X-ray observations, we examine the connection between the different emission components in these systems. Our discussion focuses on the physical nature and variability of the radio emission, as the optical and X-ray components are much better understood.

\subsection{The origin of the radio emission}

The origin of radio emission in RQ AGN remains a subject of ongoing debate, as it could likely arise from a combination of physical structures rather than a single mechanism. According to the diagnostic framework proposed by \citet{panessa2019}, the dominant process can often be inferred by combining spectral indices and morphological information across different spatial scales.

The observed emission is strongly frequency-dependent. While there is no sharp boundary between components, observations in the 1--10~GHz range are generally thought to be dominated by jet emission on parsec to kiloparsec scales \citep{blandford1979, fischer2021}, characterized by optically thin synchrotron radiation from extended structures (lobes and jets) or a flat-spectrum core representing the self-absorbed base of an adiabatic jet. In contrast, the 10--100~GHz range acts as a transition regime where the sub-parsec jet base becomes more prominent as synchrotron self-absorption decreases \citep{behar2018}. At even higher frequencies ($>100$~GHz), the jet contribution may drop, allowing the magnetized corona to dominate the millimetre/sub-millimetre emission \citep{raginski2016, delpalacio2025}. 

Our observations at 15~GHz place our sample within this transition regime. This is consistent with previous characterizations of hard X-ray selected AGN, which show predominantly compact morphologies at arcsecond scales \citep{chiaraluce2020, magno2025} and flat or inverted radio spectra between 5 and 45~GHz \citep{panessa2022}. High-resolution VLBA observations of similar samples further support this scenario, revealing that compact radio cores are nearly ubiquitous at sub-parsec scales \citep{fischer2021, shuvo2022}. These characteristics suggest that the 15~GHz emission is dominated by synchrotron self-absorption in the innermost regions, likely originating from a compact, and perhaps frustrated, jet core.

This physical nature has direct implications for the FP. Our sample consists of mainly RQ sources, and we find that the \citet{merloni2003} relation provides a consistent description of our sample, with a near-zero mean bias ($0.027$~dex). 
Our multi-epoch simultaneous monitoring reveals that variability-induced dispersion ($\sigma_{var} = 0.118$~dex) accounts for only $\sim 2\%$ of the global observed scatter in the FP. In particular, the radio variability contributes only 0.73\%, consistent with the \textit{FRAMEx} survey, which reported that radio variability on month-to-year timescales is typically modest ($\lesssim 20\%$) compared to the more dramatic fluctuations seen in X-rays \citep{fischer2021}. This suggests that the compact radio core remains a resilient feature of the accretion-ejection architecture over extended periods.
Our results show that even in sources with higher relative fluctuations, such as 4U~0517+17 or QSO~B0241+62, the impact of stochasticity remains negligible compared to the total dispersion of the plane ($0.786$~dex). 

Dramatic shifts across the FP, such as those seen in the 'changing-look' AGN Mrk~590 \citep{palit2026}, appear to be rare exceptions rather than the rule for the general AGN population. Recent population studies of changing-look events indicate that while some sources can transition from RQ to RL states on decadal timescales, potentially due to the ignition of young, intermittent jets \citep{nyland2020}, the majority of such systems do not exhibit the dramatic radio fading or flaring seen in Mrk~590 \citep{birmingham2025}. 
This suggests that the observed residuals (most notably in IGR~J21247+505 and NGC~5252) are not an artifact of non-simultaneous data or short-term variability. Instead, they likely represent intrinsic physical departures from the scaling relation, potentially reflecting differences in accretion efficiency, magnetic field geometry, or specific jet properties across the population.

While the stability of the FP points toward a robust underlying coupling between accretion and ejection, the nature of this link can be further constrained through variability. Comparing radio and X-ray fluctuations provides a powerful diagnostic: correlated variability with specific time lags can distinguish between synchrotron jets, where radio typically lags X-rays by tens of days \citep[e.g.,][]{bell2011}, and coronal models, which may exhibit different patterns depending on the heating and cooling mechanisms involved \citep{panessa2019}.

\subsection{Radio Variability Origin: Intrinsic vs. Extrinsic Effects}

To assess the nature of the observed 15 GHz variability, we evaluated the potential contribution of Refractive Interstellar Scintillation (RISS). Radio waves from compact sources are subject to scattering by electron density inhomogeneities in the Galactic interstellar medium (ISM), which can induce extrinsic flux modulations \citep{rickett1990}.

Following the formalism of \cite{walker1998}, the impact of RISS depends on the transition frequency ($\nu_0$), which defines the boundary between weak and strong scattering regimes. This frequency is a function of the Galactic coordinates (l,b) of each source. At our observing frequency of $\nu$=15 GHz, sources at low Galactic latitudes (|b|$<$10$^\circ$) generally fall into the strong scattering regime ($\nu < \nu_0$), while high-latitude sources are in the weak scattering regime ($\nu > \nu_0$).

We calculated the theoretical modulation index (m$_{theor}$), representing the maximum expected fractional variability due to RISS following relations from \cite{walker1998}. We adopted transition frequencies ($\nu_0$) from the NE2001 Galactic model, which range from approximately 4--8~GHz for high-latitude sources like NGC~4388 and NGC~5252, up to 35--50~GHz for those located near the Galactic plane, such as QSO~B0241+62.

To compare our data with these theoretical limits, we defined the observed modulation index ($m_{\rm obs}$) as the square root of the normalized excess variance ($\sigma^{2}_{NXS})$.

 At our observing frequency of 15~GHz, we find that the theoretical modulation index ($m_{\rm theor}$) predicted for RISS is generally higher than the $m_{\rm obs}$. In several cases, particularly for sources at low Galactic latitudes, the expected amplitude of scintillation could potentially account for fluctuations between 50\% and 90\% of the mean flux, which is significantly greater than the 5--20\% variability measured in our data. While this amplitude comparison might initially suggest that RISS cannot be formally ruled out as a contributor to the variability, a more definitive distinction can be made by analyzing the characteristic timescales of the fluctuations.

The characteristic timescale for RISS at 15 GHz, determined by the time required for the Earth's orbital motion to cross the diffraction pattern (the Fresnel scale) created by the interstellar medium, is expected to be remarkably short. Based on the relationship $t_r \approx 2 (\nu_0 / \nu)^{11/5}$~hours in \cite{walker1998}, the predicted fluctuations for our sample at this frequency should occur on scales of hours to a few days. However, the light curves obtained in our monitoring program exhibit coherent trends, characterized by smooth increases and decreases in flux that persist over several months or even years. This temporal behavior is physically inconsistent with the fast, stochastic ``flicker'' that RISS would produce at 15~GHz. While RISS timescales can extend to months or years at lower frequencies (sub-GHz) due to the steep frequency dependence of the scattering regime, this is not the case for our observations. Consequently, the long-term coherence of the recorded variations, combined with the fact that sources like NGC~5252 exhibit modulation indices that exceed even the most pessimistic RISS predictions, strongly indicates that the observed 15~GHz variability is primarily intrinsic to the AGN.

\subsection{Multiwavelength variability: Hierarchy and Constraints}
\label{sec:variability_hierarchy}

The coordination of high-cadence, multi-wavelength monitoring campaigns presents significant observational challenges, particularly when seeking to synchronize different facilities. Despite these constraints, dedicated studies of individual sources have been instrumental in identifying the different physical regimes that can govern the RQ population. On one hand, some objects exhibit a high degree of synchronization across bands, suggesting a compact and direct connection between the accretion flow and the radio-emitting plasma. A prime example is Mrk 110, where daily radio variability at milli-arcsecond scales points toward a very compact core \citep{panessa2022b}. Similarly, the tight coupling observed in NGC 2992 \citep{fernandez2022} and the fluctuations reported in other Seyferts by \cite{chen2022} reinforce the scenario where the radio emission tracks the energetic changes of the corona on timescales of a few weeks.

On the other hand, several studies have highlighted cases where the radio emission appears physically decoupled from the rapid fluctuations of the inner engine. In these instances, the radio flux remains remarkably stable even during dramatic X-ray or optical transitions. This was notably observed in NGC 4388, where the radio emission is likely dominated by persistent structures such as shocks in AGN-driven winds or extended outflows \citep{sargent2024, sargent2025}. A similar decoupling was found in the monitoring of the weak-line quasar SDSSJ1539+3954 \citep{chhipa2026}, where the radio emission failed to respond to a major X-ray drop. These diverse behaviors suggest that the nature of the radio-emitting region can vary significantly even among sources with similar classification.

The sources in our sample exhibit a diverse range of variability patterns when considering the three energy bands simultaneously. 
Two sources show fluctuations across the entire spectrum, with the largest amplitudes observed in X-rays and the most moderate in the radio band (LEDA 168563, and MCG+08-11-11). Qualitatively, three other sources could follow this trend (4U 0517+17, Mkn 6, and RX J2135.9+4728) but variations in the radio band above 3$\sigma$ cannot be confirmed.
Three sources (IGR\,J00333+6122, 2E 1853.7+1534, and IGR\,J21247+5058) display variations in X-rays and could show hints of variations in the radio (which cannot be confirmed) but remain remarkably stable in the optical. For these type 1 objects, this suggests a potential decoupling between the optical and higher-energy emission, or simply that the r-band is too dominated by the host galaxy to detect small changes. Further monitoring in cleaner bands would be required to confirm this.
Among our sample, only Mkn\,3 and IGR\,J23308+7120 (both classified as Seyfert 2) show no significant variability in any band. 
However, Mkn\,3 has been reported to be variable in X-rays in previous studies \citep[e.g.,][]{guainazzi2012, lore2015}; therefore, it is possible that the source did not exhibit variability during our observation period, or that the uncertainties are too large to preclude its detection.
The remaining four sources present distinct behaviors: QSO B0241+62  exhibits fluctuations across all three bands, but with a unique amplitude hierarchy where the X-ray variability is the strongest, followed by the radio, and finally the optical; this contrasts with other sources of the sample where optical variations typically exceed those in the radio. Furthermore, NGC 4388 varies exclusively in the radio regime; NGC 5252 shows comparable variability amplitudes in both radio and X-rays; and SWIFT J2127.4+5654, the only NLSy1 in our sample, shows variability restricted to the X-ray band, with no hints of radio varability. Interestingly, the sources showing synchronized variations across all wavebands are consistently located within the RQ regime in both the X-ray and optical diagnostic planes. Conversely, those exhibiting no optical variations or other non-standard patterns tend to occupy the region of the diagram characterized by 'radio intermediate' (Fig. \ref{fig:radio_loudness_diagnostic}).

The diversity of variability patterns and the general observed hierarchy in amplitudes (X-rays $>$ Optical $>$ Radio; Fig. \ref{fig:mexican_hat_plot}) point toward a complex flow of energy from the innermost regions to the more extended structures. This trend is remarkably consistent with the X-ray and optical variations being strongly correlated, a fact usually attributed to X-ray reprocessing, where the corona illuminates the disk, which then re-emits the energy at longer wavelengths \citep{arevalo2009,breedt2009,lira2015}.
Our data suggests that in these sources, the chain could extend further out to the radio-emitting region, 
likely a magnetized corona or a
very compact jet base that responds to the same energetic drivers
\citep{laor2008,panessa2019}.

However, the decoupling observed in some sources—where X-ray and radio variations persist despite a stable optical flux—could indicate a shift in the dominant emission mechanism. As sources approach the RL regime, radio emission may become increasingly independent of the accretion disk's thermal state, driven by episodic jet components or magnetic reconnection in the corona \citep{merloni2001, king2013}. This effectively bypasses the slower viscous timescales of the disk and explains the lack of evident X-ray-to-optical reprocessing, as the energy is preferentially channeled into non-thermal structures rather than being re-emitted by the disk. While alternative models suggest parsec-scale shocks from AGN winds \citep{yamada2024}, the lack of ultrafast outflows in our sample makes this less likely. Alternatively, it remains possible that the optical emission is simply too dominated by the host galaxy to detect the low-amplitude variations expected from reprocessing.

To distinguish among the various physical scenarios discussed above, future multi-wavelength monitoring with a daily-to-weekly cadence would be essential. Such high-cadence data is required to resolve short-term variability and determine the physical time lags between the different emission components, which is key to mapping the connection between the accretion disk, the corona, and the jet.

\section{Conclusions}
\label{sec:conclusions}

In this work, we have characterized the multi-wavelength variability of a sample of 14 hard X-ray selected AGN using 15~GHz, optical ($g$ and $r$), and 2--10~keV X-ray observations. Our main results and their physical implications are summarized as follows:

\begin{enumerate}
    \item Significant variability was detected in 86\% (12/14) of the sample. Across individual bands, 11 sources (79\%) were found to vary in X-rays, 8 (57\%) in the optical, and 6 (43\%) in the radio. Only two sources, Mkn~3 and IGR~J23308+7120, remained stable across all monitored frequencies during the two-year period.
    
    \item The $F_{\rm var}$ reveals a clear hierarchy across the electromagnetic spectrum, with median values of $\sim$30\% in X-rays (range: 11--67\%), 19\% and 8.5\% in the $g$ and $r$ bands (2--33\% and 0.2--24\%, respectively), and 10\% in the radio (4--23\%). The Mexican Hat analysis consistently shows higher power at the 200-day scale compared to 70 days, characteristic of a red-noise stochastic process.
    
    \item According to the optical proxy ($\log R_{\rm opt} > 1$), 8\% of the sample is classified as RL. However, when using the X-ray definition ($\log R_X > -2$), the RL fraction drops to 0\%, with the majority of sources (57\%) being strictly RQ and 43\% classified as radio-intermediate.
    
    \item Our analysis confirms that the FP remains robust against monthly variability. By comparing time-averaged results with strictly simultaneous data, we found that stochastic shifts within the plane account for only $\sim$3\% of the total observed scatter ($0.786$~dex), indicating that non-simultaneity is not a dominant source of dispersion in this scaling relation.
\end{enumerate}

At 15 GHz, the radio emission is expected to be core-dominated, likely originating from a compact jet base or a magnetized corona. The observed diversity in variability patterns—ranging from highly synchronized disk-corona-jet fluctuations to instances of decoupling—reflects the intrinsic physical complexity of these AGN cores. This diversity could be linked to differences in accretion efficiency or magnetic field configurations. To further distinguish between specific emission mechanisms, such as synchrotron jets versus coronal models, future multi-wavelength monitoring campaigns with higher cadence (daily to weekly) are essential to resolve the physical time lags between spectral components.

\section*{Acknowledgements}

We thank the referee for their constructive comments, which helped improve the quality of this manuscript.
LHG acknowledges financial support from ANID program FONDECYT Iniciaci\'on 11241477. FP acknowledges financial support from the Bando Ricerca Fondamentale INAF and "Programma di Ricerca Fondamentale INAF 2023 and 2024. DRAW was supported by the Oxford Centre for Astrophysical Surveys, which is funded through generous support from the Hintze Family Charitable Foundation.
PA acknowledges financial support from ANID FONDECYT Regular 1241422,  Millennium Science Initiative Program NCN$2023\_002$ (TITANS) and CAV CIDI N. 21 of the U. de Valparaíso, Chile. 
A.M.M.A. acknowledges support from ANID BASAL Project CMM (FB210005), and ANID Millennium Science Initiative MAS (ICN12\_009 and AIM23-0001).
The research leading to these results has received funding from the
European Union’s Horizon 2020 Programme under the AHEAD2020 project (grant
agreement n. 871158).
We acknowledge the use of public data from the \emph{Swift} data archive.
The ZTF forced-photometry service was funded under the Heising-Simons Foundation grant \#12540303 (PI: Graham).
We thank the Mullard Radio Astronomy Observatory staff for scheduling and carrying out the AMI observations. 
AMI is supported by the Universities of Cambridge and Oxford, and acknowledges support from the European Research Council under grant ERC-2012-StG-307215 LODESTONE.


\bibliographystyle{aa}
\bibliography{GB_bibliography} 

\newpage


\appendix

\section{Notes on individual sources}
\label{appendix:individualsources}

\subsection{IGR\,J00333+6122}

This source is classified as a Seyfert 1.5 galaxy based on its optical spectrum \citep{masetti2009}. In the radio band, the 1.4~GHz NVSS image shows a compact, unresolved morphology at 45$''$ resolution \citep{panessa2015}. This compact nature is maintained at higher resolutions ($\sim 1''$) with the JVLA at 22 and 45~GHz, where the source remains unresolved \citep{chiaraluce2020}. Consequently, it is classified as a core-dominated source (type A) by \citet{panessa2022}, based on high-sensitivity (few $\mu$Jy~beam$^{-1}$) (sub)-kpc JVLA observations between 5 and 45~GHz. Regarding its X-ray properties, the source exhibits a variability index of 1.53 in the 54-month Palermo Swift-BAT hard X-ray catalogue \citep{cusumano2010}. More recently, the \emph{Swift}/BAT 157-month survey \citep{swift157} shows slight flux variations in the 14--195~keV light curve.

\subsection{QSO\,B0241+62}

This object is classified as a radio-loud Type 1 Seyfert galaxy \citep{apparao1978}. In the optical band, photometry has revealed small variations and potential outbursts on timescales of months to years \citep[see][and references therein]{bozyan1990}. The radio morphology, observed with the VLA and VLBA, has been interpreted as a core-jet structure \citep{preston1985, lister1998, charlot2010}, specifically identified as a one-sided jet (type B) by \citet{panessa2022}. Flux density measurements in the range of 700--900~mJy, combined with a flat spectral index, suggest a jet closely aligned with the line-of-sight \citep{chiaraluce2020}. Furthermore, \citet{lister1994} reported a radio variability index of 0.16 at 6~cm between 1982 and 1984. High-energy observations also show significant activity; the IPC instrument on board the Einstein Observatory revealed X-ray flux variability by more than a factor of two in the 0.5--4.5~keV band over a six-month period \citep{mereghetti1985}. Consistently, the source is included in the \emph{Swift}/BAT 157-month survey \citep{swift157}, where its 14--195~keV light curve exhibits continued slight variations.

\subsection{LEDA\,168563 }

This source was classified as a Seyfert 1 galaxy by \citet{winter2010}. The first radio study of this object was conducted by \citet{panessa2015}, who reported a flux density of $\sim$15~mJy at 1.4~GHz with NVSS. Subsequent radio maps by \citet{chiaraluce2020} revealed a compact morphology that appears slightly resolved at 22~GHz but remains unresolved at 45~GHz. Consequently, it is classified as a core-dominated source (type A) by \citet{panessa2022}. Regarding its high-energy emission, \citet{liebmann2018} reported significant long-term X-ray variability over a six-year period, comparing \emph{Suzaku} and \emph{XMM-Newton} observations. During this time, the source exhibited a decrease in the soft excess and a flattening of the power-law component, while the hard X-ray Compton hump flux relatively increased; these variations were interpreted within an ionized relativistic reflection scenario. Finally, the source is included in the \emph{Swift}/BAT 157-month survey \citep{swift157}, where its 14--195~keV light curve shows continued slight variations.

\subsection{4U\,0517+17}

This source is optically classified as a Seyfert 1.5 galaxy \citep{remillard1993}. In the radio band, \citet{panessa2015} derived a flux of $\sim$6.1~mJy with NVSS (45$''$ resolution), identifying a slightly resolved morphology. Subsequent JVLA observations by \citet{chiaraluce2020} revealed a compact structure that appears slightly resolved at 22~GHz and unresolved at 45~GHz. This radio emission has been interpreted as originating from either the optically-thick base of a jet or from the corona, and the source is classified as core-dominated (type A) by \citet{panessa2022}. Regarding its X-ray variability, \citet{soldi2014} analyzed the \emph{Swift}/BAT 58-month survey, reporting a variability estimator of 17$^{+6}_{-8}$ per cent in the 14--195~keV energy band and 3$\pm$2 per cent in the 14--24~keV range.

\subsection{MCG+08-11-11}

This galaxy has been extensively studied in the literature; here we highlight the most recent variability studies and those reporting significant spectral variations. Optically, it is classified as a Seyfert 1.5 galaxy \citep{fausnaugh2017}. Photometric monitoring has revealed variations of up to 0.5~mag on monthly timescales, with the largest amplitudes observed in the bluer filters \citep{webb2000}. More recently, \citet{lu2019} reported a fractional variability of $F_{var} = 9.72 \pm 0.52\%$ in the context of black hole mass and reverberation mapping relations. Notably, \citet{fausnaugh2017} identified this source as having the largest optical variations among their reverberation mapping sample, deriving a black hole mass of $\sim 2.8 \times 10^7\,M_{\odot}$. 

In the radio band, the source exhibits an elongated morphology in the N--S direction \citep{chiaraluce2020}, and it is classified as either a one-sided jet (type B) or a triple structure (type C) by \citet{panessa2022}. Regarding its high-energy emission, the source has shown significant X-ray variability since the early stages of the \emph{Swift} mission, with variations of 59\% on daily timescales and 26\% over 20 days \citep{beckmann2007}. Analysis of the \emph{Swift}/BAT 58-month survey yields a variability estimator of $33^{+5}_{-4}$\% in the 14--195~keV band \citep{soldi2014}. Finally, using simultaneous data from NOEMA (mm) and NuSTAR (X-rays), \citet{petrucci2023} reported the detection of synchronized intra-day variability on timescales of $\sim$5~hours, suggesting that the millimetre emission originates from a very compact region ($\lesssim 1300\,R_g$) likely associated with the X-ray corona or a weak jet base.

\subsection{Mkn\,3}

This source was optically classified as a Seyfert 2 galaxy \citep{Khachikian1974}, although broad lines have been detected in polarized light \citep{miller1990}, revealing its hidden broad-line region. In the radio band, observations at different frequencies have revealed a double nucleus structure \citep{Ulvestad1984}, and it is classified as a one-sided jet (type B) by \citet{panessa2022}. Regarding its X-ray properties, the source was historically characterized as Compton-thick \citep{bassani1999} but was later reclassified as heavily obscured \citep{guainazzi2016}, with a column density of $N_H \approx [0.8-1.1] \times 10^{24}$~cm$^{-2}$. Significant X-ray variations have been reported on timescales of months and years in the nuclear continuum \citep{guainazzi2012, lore2015}, including a notable occultation event lasting approximately one month in 2014 \citep{guainazzi2016}. Furthermore, analysis of the \emph{Swift}/BAT 58-month survey yields a variability estimator of $35^{+11}_{-9}$ per cent in the 14--195~keV band \citep{soldi2014}, and the source continues to show strong flux variations in the more recent \emph{Swift}/BAT 157-month survey catalog \citep{swift157}.
 
\subsection{Mkn\,6}

This galaxy is optically classified as a Seyfert 1.5 \citep{Osterbrock1976}. Since the 1970s, the source has exhibited strong optical variations in both the continuum and spectral lines on timescales ranging from weeks to months \citep[e.g.,][]{rosenblatt1989, eracleous1993, Afanasiev2014}. In the radio band, \text{Mkn~6} displays a very characteristic morphology that has been interpreted as either jet precession or the interaction of the jet with the ambient interstellar medium \citep{kharb2006, chiaraluce2020}. Due to this structural complexity, it is classified as a one-sided jet (type B), triple (type C), or jet+complex (type D) source by \citet{panessa2022}. This complexity extends to the X-ray regime, where the source has shown variations in the absorbing column density over timescales of years; these changes have been modeled using a partial covering scenario or a warm absorber \citep{Immler2003, Schurch2006}. Furthermore, analysis of the \emph{Swift}/BAT 58-month survey yields a variability estimator of $18^{+5}_{-6}$ per cent in the 14--195~keV energy band \citep{soldi2014}, with the 157-month survey catalog \citep{swift157} confirming continued slight variations in the same hard X-ray range.

\subsection{NGC\,4388}

This object is a well-known Seyfert 2 galaxy \citep{Phillips1982} located in the Virgo cluster, notable for its ionization cones \citep{ayani1989} and the detection of a water maser \citep{kuo2011}. In the optical and near-infrared (NIR), long-term monitoring over a decade has revealed significant variability; for instance, \citet{Dahmer2023} reported an 88\% decline in warm dust contribution and a 61\% decrease in the [Ca~{\sc viii}] coronal line, while the Br$\gamma$ emission declined by 35\%. Regarding its radio properties, the source exhibits a double-peaked jet structure, with the northern component typically identified as the nucleus \citep{hummel1991, chiaraluce2020}. It is classified as a core-dominated one-sided jet (type B) by \citet{panessa2022}. The X-ray emission shows high complexity and significant variations in the absorbing column density, with changes of 50\% observed over 12-month timescales \citep{risaliti2002}. Furthermore, combined \emph{INTEGRAL} and \emph{Swift} observations from 2003 to 2009 show flux and spectral slope variations on monthly timescales \citep{Fedorova2011}. Finally, \citet{soldi2014} reported a variability estimator of 26$\pm$2\% in the 14--195~keV band, and the source continues to exhibit strong variations in the 157-month \emph{Swift}/BAT survey \citep{swift157}.

\subsection{NGC\,5252}

This galaxy is optically classified as a Seyfert 1.9 \citep{Osterbrock1993} and is well known for displaying large-scale ionization cones in both the optical and X-ray bands \citep{wang2024}. Regarding its near-infrared properties, \citet{quillen2000} analyzed \emph{Hubble Space Telescope} observations at 1.6~$\mu$m but reported no significant variations on monthly timescales. In the radio band, VLA observations at arcsecond resolution reveal an unresolved core—coincident with the optical nucleus—surrounded by weak, resolved jet-like extensions to the north and south \citep{wilson1994}. Consequently, the source is classified as a core-dominated morphology (type A) by \citet{panessa2022}. The X-ray emission is characterized by significant spectral and flux variability; \citet{risaliti2002} reported changes in the column density of 36\% over a four-year period, a behavior later confirmed by \citet{laha2020} using \emph{XMM-Newton}, \emph{Chandra}, and \emph{Suzaku} data. Additionally, an off-nuclear ultraluminous X-ray source (ULX; CXO J133815.6+043255) was discovered 22$''$ from the nucleus \citep{kim2015}. Finally, the source is included in the \emph{Swift}/BAT 157-month survey \citep{swift157}, where its 14--195~keV light curve exhibits strong variations.

\subsection{2E\,1853.7+1534 }

This source was classified as a Seyfert 1 galaxy by \citet{masetti2006}. In the radio band, it exhibits a compact morphology and is classified as a core-dominated source (type A) by \citet{panessa2022}. Regarding its high-energy properties, a variability index of 1.44 was reported based on the first 54 months of the \emph{Swift} mission \citep{cusumano2010}, indicating a lack of strong hard X-ray variations during that period.

\subsection{IGR\,J21247+5058}

The optical spectrum of this source exhibits a broad H$\alpha$ line, leading to its classification as a Seyfert 1.9 galaxy \citep{masetti2004}. This object is associated with the radio galaxy 4C~50.55 \citep{ribo2004}, and VLA observations allow for its classification as a Fanaroff-Riley type II (FRII) Broad Line Radio Galaxy \citep{molina2007}. In the X-ray band, the source shows significant variability across different missions and energy ranges. Analysis of \emph{INTEGRAL} data in the 20--40~keV band revealed a fractional variability of $0.270 \pm 0.039$ \citep{telezhinsky2010}, while more recent \emph{NuSTAR} observations show a normalized excess variance of 0.054 \citep{papadakis2024}. Consistently, the source is included in the \emph{Swift}/BAT 157-month survey \citep{swift157}, where its 14--195~keV light curve exhibits strong variations.

\subsection{SWIFT\,J2127.4+5654}

This source is optically classified as a narrow-line Seyfert 1 (NLSy1) galaxy \citep{halpern2006, malizia2008}. In the radio band, observations at 1.4~GHz show a single, slightly resolved morphology \citep{panessa2015}. The X-ray emission of this object has been extensively studied; notably, \citet{miniutti2009} used \emph{Suzaku} data to detect a relativistic iron line, deriving an intermediate black hole spin ($a \sim 0.6$) and suggesting that the accretion may be driven by chaotic episodes or magnetic extraction of rotational energy. Regarding its variability, the source is included in the \emph{Swift}/BAT 157-month survey \citep{swift157}, where its light curve exhibits slight variations in the 14--195~keV band. On shorter timescales, \citet{gozalezmartin2012} analyzed the power spectrum density (PSD) from \emph{XMM-Newton} data, finding significant variations across soft, hard, and total energy bands. Spectral variability in the same data has been attributed to a partial eclipse of the X-ray source by an intervening low-ionization or cold absorbing structure \citep{sanfrutos2013}. Furthermore, multi-epoch \emph{NuSTAR} observations reveal strong fractional variability ($F_{var}$) on daily timescales \citep{rani2017}. Finally, combined \emph{NuSTAR} and \emph{XMM-Newton} analyses suggest that these variations are primarily driven by changes in the power-law normalization and ionized reflection, coupled with variable absorption at lower energies \citep{kammoun2018}.

\subsection{RX\,J2135.9+4728}

This source was classified as a Seyfert 1 galaxy by \citet{burenin2008}. In the radio band, its 1.4~GHz morphology was studied by \citet{panessa2015}, who reported it as a single, resolved source. Regarding its high-energy emission, the object is included in the \emph{Swift}/BAT 157-month survey catalog \citep{swift157}, where its 14--195~keV light curve exhibits slight flux variations.

\subsection{IGR\,J23308+7120}

This object was classified as a likely Seyfert 2 galaxy by \citet{masetti2008}. In the radio band, the source exhibits a compact morphology, as noted by \citet{smith2016}, and is specifically classified as a core-dominated source (type A) by \citet{panessa2022}. Regarding its high-energy emission, the source is included in the \emph{Swift}/BAT 157-month survey catalog \citep{swift157}, although its 14--195~keV light curve shows no significant variations during this period.

\section{Photometric error characterization and variability classification of ZTF data}
\label{appendix:variability}

\begin{figure}
    \centering
    \includegraphics[width=\linewidth]{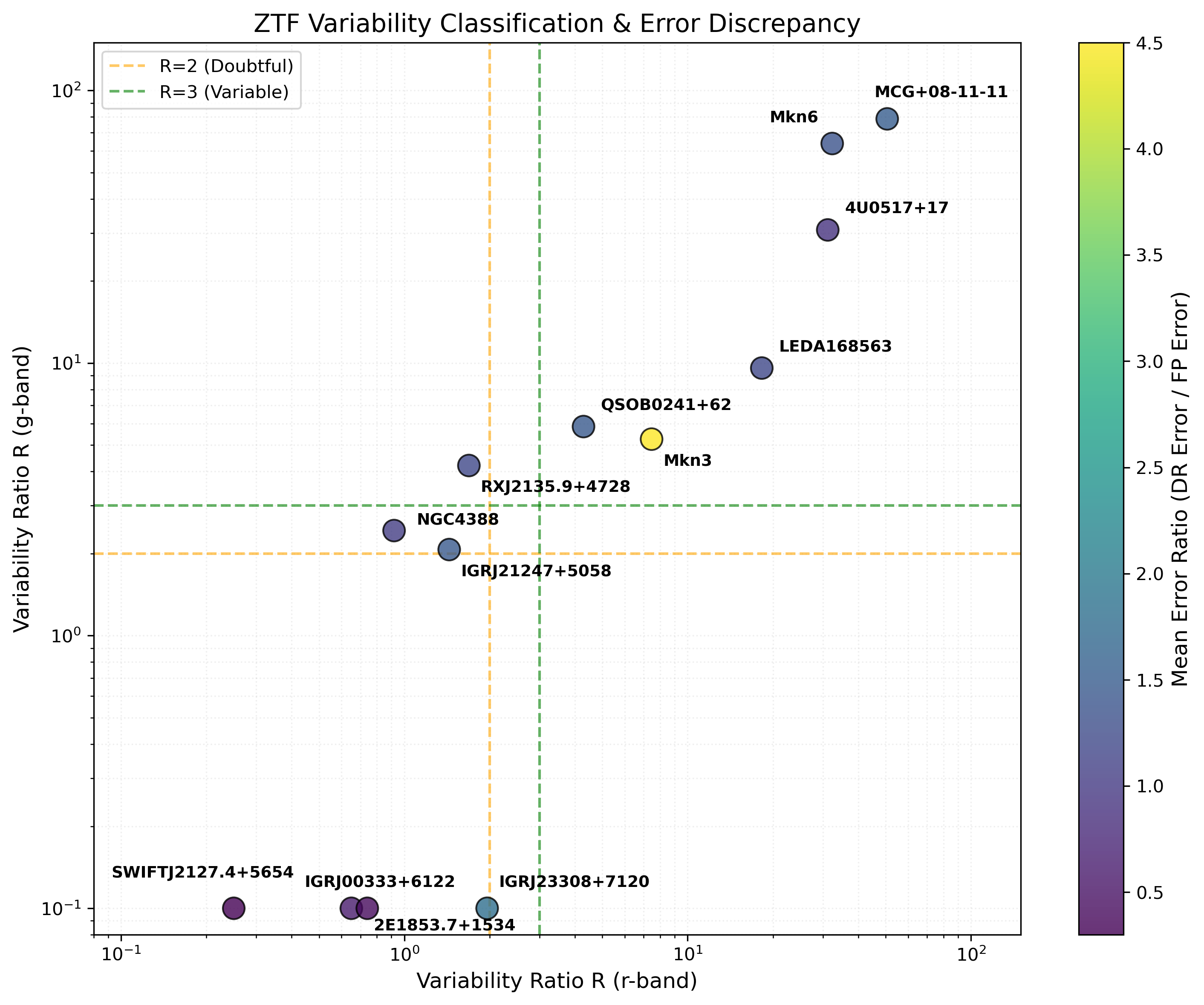}
    \caption{Variability ratio $R$ in $g$ vs $r$ bands. Dotted lines indicate the adopted thresholds for Non-variable ($R < 2$) and Variable ($R > 3$) sources. The color bar represents the Error Ratio (DR error / FP error). Mkn3 (yellow point) is clearly identified as an outlier with high error discrepancy, justifying its exclusion. Sources in the "Doubtful" region (g band of IGR J21247+5058 and NGC 4388) had those specific bands excluded to maintain a conservative variability sample.}
    \label{fig:variability_analysis}
\end{figure}

The reliability of AGN variability studies is inherently tied to the accuracy of the reported measurement uncertainties. While ZTF forced photometry is optimized for flux measurement at fixed coordinates, effectively mitigating host galaxy contamination, it can occasionally underestimate the noise floor in regions of high surface brightness or complex backgrounds \citep{arevalo2026}.

Indeed, \cite{masci2023} report corrections that need to be applied to the photometric uncertainties under specific conditions. Here, we find that even after applying these corrections, a small number of sources not expected to show optical nuclear variability 
(e.g., Seyfert 2 galaxies) are classified as variable, suggesting that the uncertainties may still be underestimated. Consequently, we performed a multi-step validation to distinguish physical variability from systematic artifacts.

We compared the photometric uncertainties obtained from the ZTF forced photometry and the ZTF Data Release (DR) 19 for all sources. DR light curves were retrieved from IRSA\footnote{https://irsa.ipac.caltech.edu/}. Since DR photometry is performed on science images where the host galaxy is present, its errors often provide a more realistic representation of the total noise budget at the source position. The two datasets were merged using a tolerance of one day to enable a direct comparison of the uncertainties. We note that only the uncertainties were compared, and not the photometric measurements themselves, since the latter are derived from science and difference images and can exhibit large scatter for non-variable sources, making a direct comparison unreliable. 

For each source and filter, we calculated the \textit{Error Ratio}, $E_{ratio}$, as the ratio of the error from the DR and the forced photometry, a metric where values near unity indicate consistent noise modeling.

Then, to isolate genuine astrophysical signals, we define the variability ratio $R = S_{\rm obs}/S_{\rm err}$, where $S_{\rm obs}$ is the observed light curve dispersion (estimated through the Median Absolute Deviation) and $S_{\rm err}$ is the expected noise dispersion, estimated as the median of the photometric uncertainties.

The selection of thresholds for $R$ was calibrated using the Seyfert 2 objects in the sample (NGC 4388 and IGR J23308+7120) to serve as empirical `noise baselines'; their lack of optical variability allows us to map the distribution of $R$ values produced by pure photometric noise. Based on this distribution, we established the following classification framework:

\begin{itemize}
    \item Significant Variability ($R \ge 3$): A $3\sigma$ threshold ensures that the observed signal is dominated by intrinsic AGN fluctuations. This level of significance is required for a band to be included in the calculation of physical variability parameters.
    \item Doubtful/Marginal Variability ($2 \le R < 3$): In this regime, the signal-to-noise ratio is insufficient to rule out subtle systematic effects or slight error underestimations. 
    \item Non-Variable ($R < 2$): The dispersion is fully consistent with the noise floor defined by the Seyfert 2 sources.
\end{itemize}

Table \ref{tab:variability_flags_final} and Fig. \ref{fig:variability_analysis} show the results of this analysis. 
The majority of the variable AGN sample exhibits $E_{ratio}$ $\approx 1$, confirming the robustness of the forced photometry uncertainties for these sources. However, Mkn\,3 displays a severe discrepancy (Ratio$_r = 5.4$, Ratio$_g = 3.6$), indicating that the forced photometry errors are underestimated by a factor of 4--5. This source is yellow-colored in Fig. \ref{fig:variability_analysis} for this reason. Considering that Mkn\,3 is a Seyfert 2 galaxy—physically expected to be stable in the optical, we conclude that its high variability ratio $R$ is a spurious result of this error underestimation. To ensure the statistical purity of the sample, ZTF data of Mkn\,3 was formally excluded from further analysis. This is also the reason why we didn't use it for the `noise baseline'.

On the other hand, we identified sources where one band showed significant variability while the other remained in the `doubtful' range. For IGR\,J21247+5058 ($R_g=2.07$) and NGC\,4388 ($R_g=2.43$), the $g$-band data were discarded, to not introduce bias in the variability analysis. This cleaning process ensures that any source classified as `Variable' in our final selection (Table \ref{tab:variability_flags_final}) possesses at least one band where the variability is unequivocally real.

\begin{table*}
\centering
\caption{ZTF data variability classification and error validation. $R$ is the variability ratio ($S_{\rm obs}/S_{\rm err}$), and $E_{ratio}$ is the error discrepancy between DR and forced photometry data. Sources are classified as Variable, Non-var, or Excluded according to the combination of these two ratios (see text for details).}
\label{tab:variability_flags_final}
\begin{tabular}{lccccc}
\hline
Name & $R_r$ ($E_{ratio,r}$) & Flag$_r$ & $R_g$ ($E_{ratio,g}$) & Flag$_g$ & Final Class  \\
\hline
IGR\,J00333+6122 & 0.74 (0.4) & Non-var & -- & -- & Non-var  \\
QSO\,B0241+62 & 4.29 (1.5) & Variable & 5.85 (1.4) & Variable & Variable  \\
LEDA\,168563 & 18.27 (1.2) & Variable & 9.59 (1.2) & Variable & Variable  \\
4U\,0517+17 & 31.23 (0.9) & Variable & 30.80 (0.9) & Variable & Variable  \\
MCG+08-11-11 & 50.63 (1.4) & Variable & 78.69 (1.6) & Variable & Variable  \\
Mkn\,3 & 7.46 (5.4) & -- & 5.26 (3.6) & -- & Excluded  \\
Mkn\,6 & 32.38 (1.0) & Variable & 63.90 (1.7) & Variable & Variable  \\
NGC\,4388 & 0.92 (0.9) & Non-var & 2.43 (1.2) & (Excluded) & Non-var  \\
2E\,1853.7+1534 & 0.65 (0.6) & Non-var & -- & -- & Non-var  \\
IGR\,J21247+5058 & 1.44 (1.1) & Non-var & 2.07 (1.8) & (Excluded) & Non-var  \\
SWIFT\,J2127.4 & 0.25 (0.3) & Non-var & -- & -- & Non-var  \\
RX\,J2135.9+4728 & 1.69 (1.1) & Non-var & 4.21 (1.3) & Variable & Variable  \\
IGR\,J23308+7120 & 1.96 (1.8) & Non-var & -- & -- & Non-var  \\
\hline
\end{tabular}
\end{table*}

\section{Plots multiwavelength light curves}
\label{appendix:plots}

This section presents the multi-frequency light curves for the AGN sample monitored in this work. Each figure displays three vertically aligned panels covering the radio emission with AMI, observed optical photometry from ZTF, and unabsorbed X-ray flux from \textit{Swift}. The temporal evolution is shown in Modified Julian Date (MJD), with all error bars representing 1$\sigma$ uncertainties. The light curves are presented in Fig. \ref{fig:lcs_full}, and details on the data reduction is shown in Section \ref{sec:observations}.

\begin{figure*}
    \centering
    \vspace{-1cm}
    \includegraphics[width=0.5\textwidth]{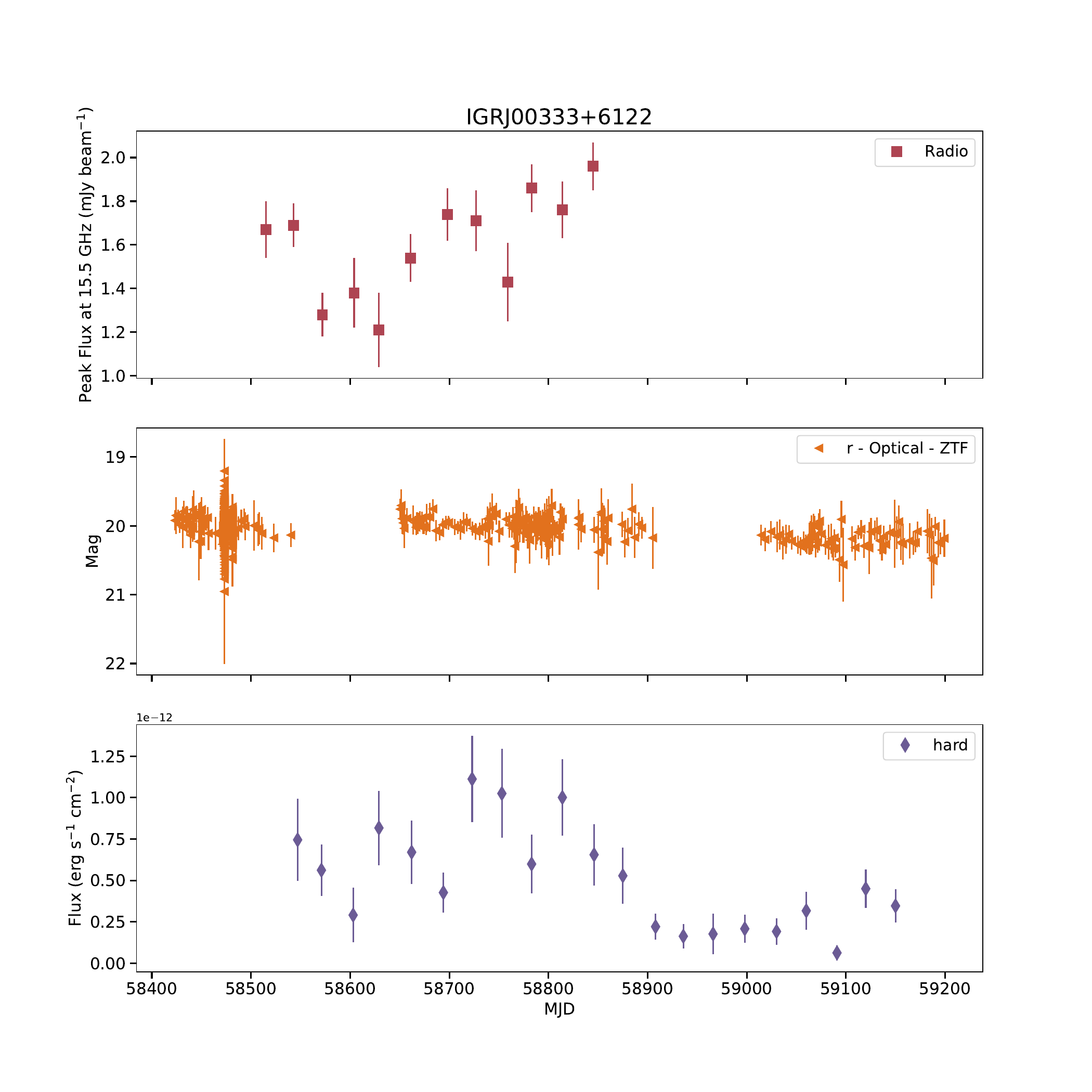}%
    \includegraphics[width=0.5\textwidth]{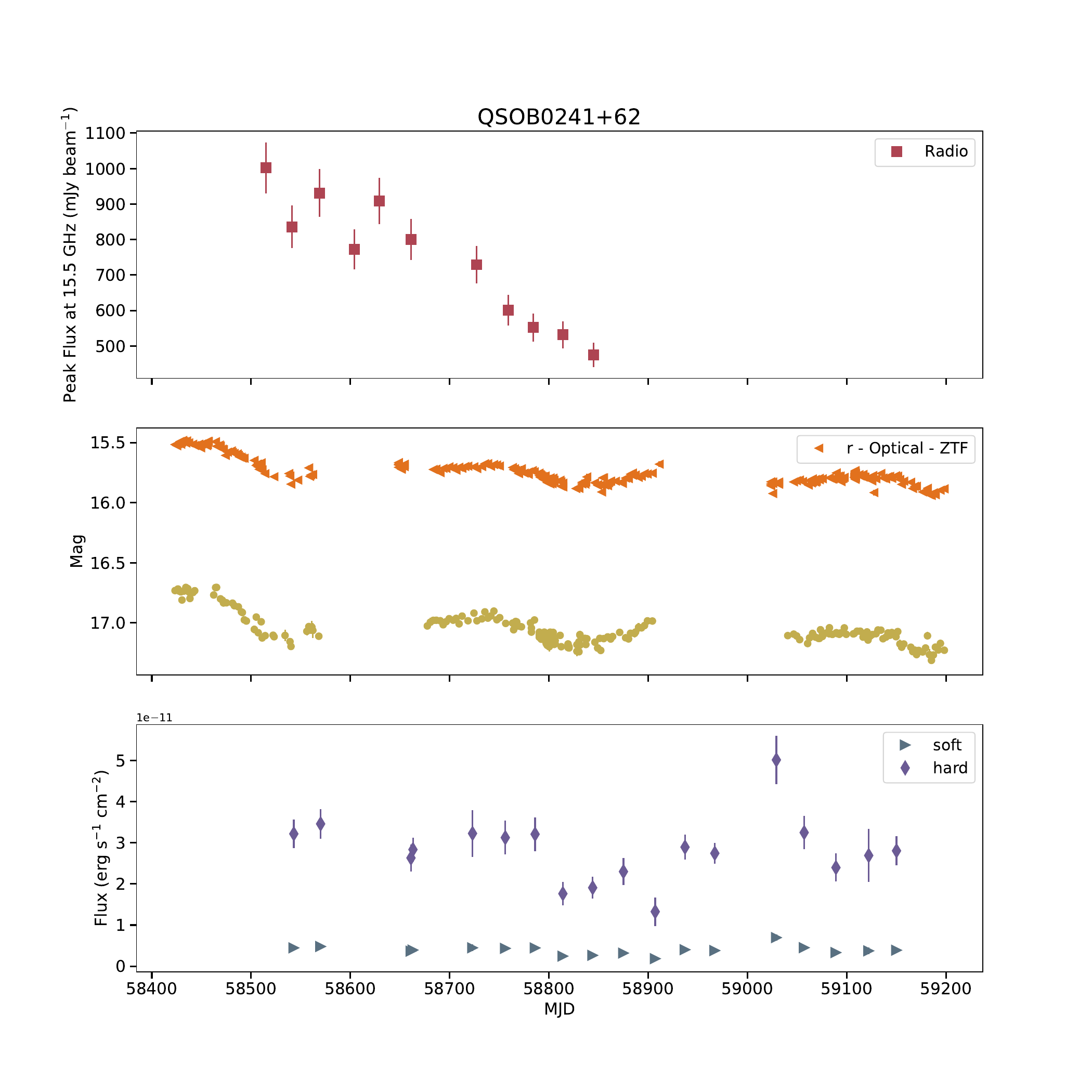}
    \vspace{-1.0cm} 
    
    \includegraphics[width=0.5\textwidth]{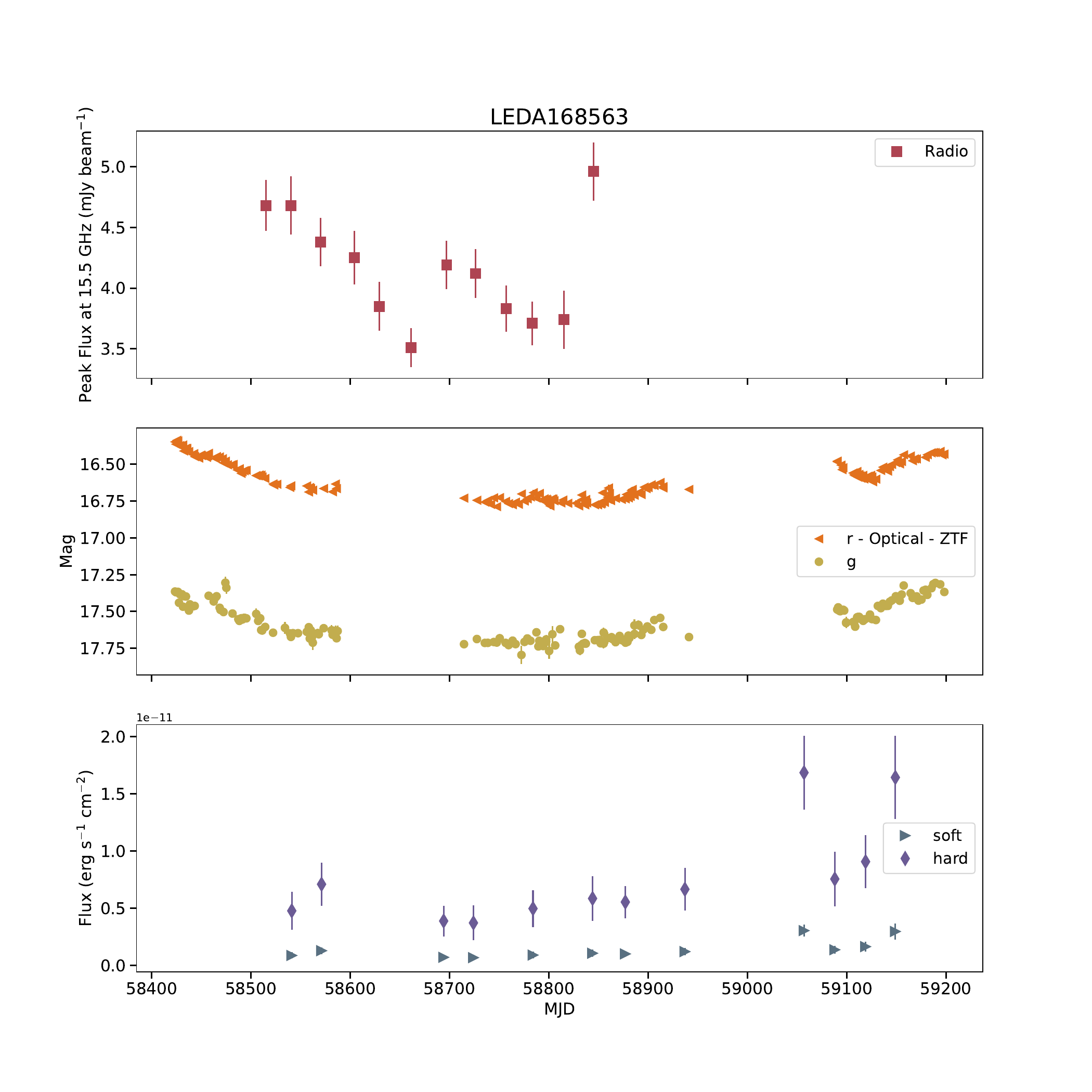}%
    \includegraphics[width=0.5\textwidth]{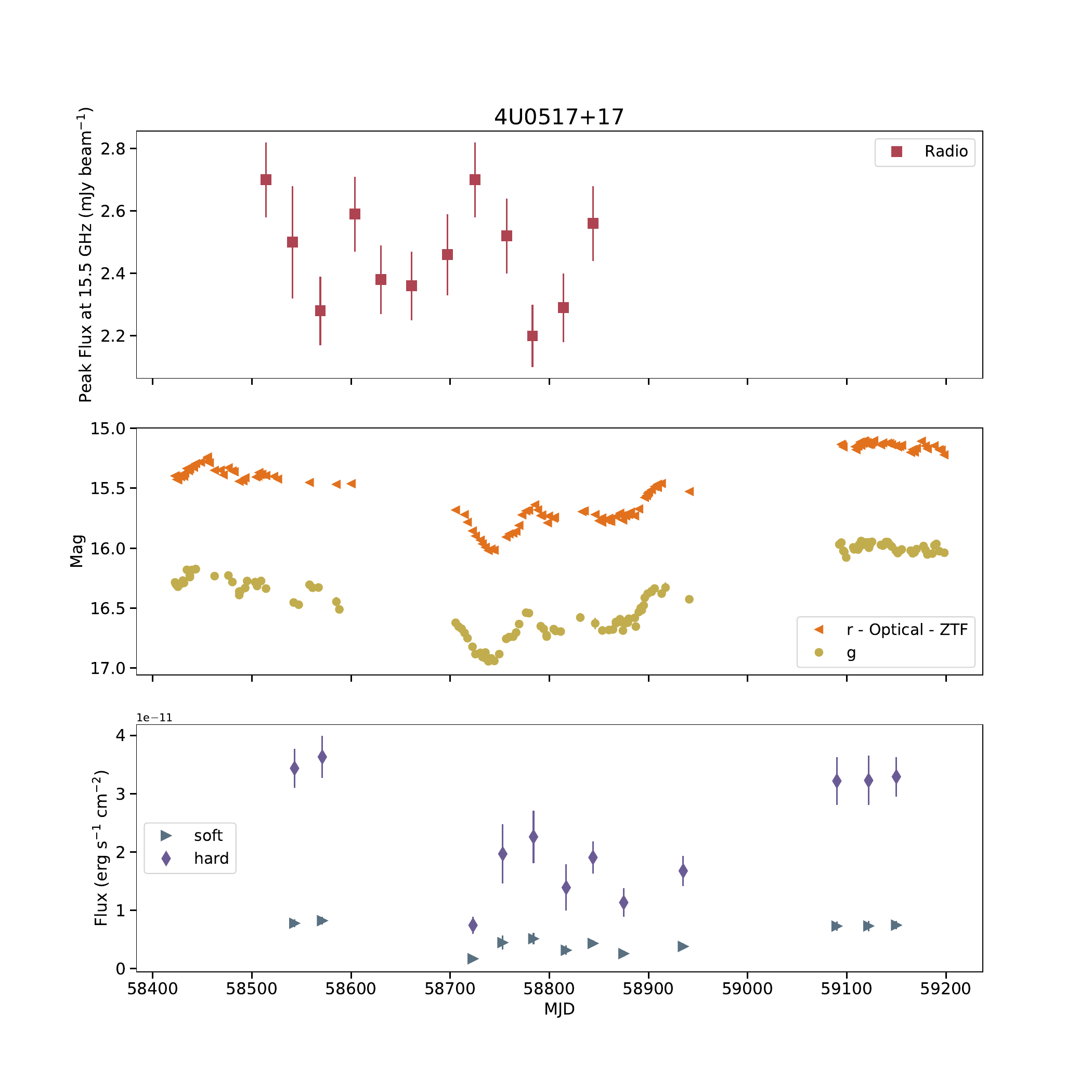}
    \vspace{-1.0cm}
    
    \includegraphics[width=0.5\textwidth]{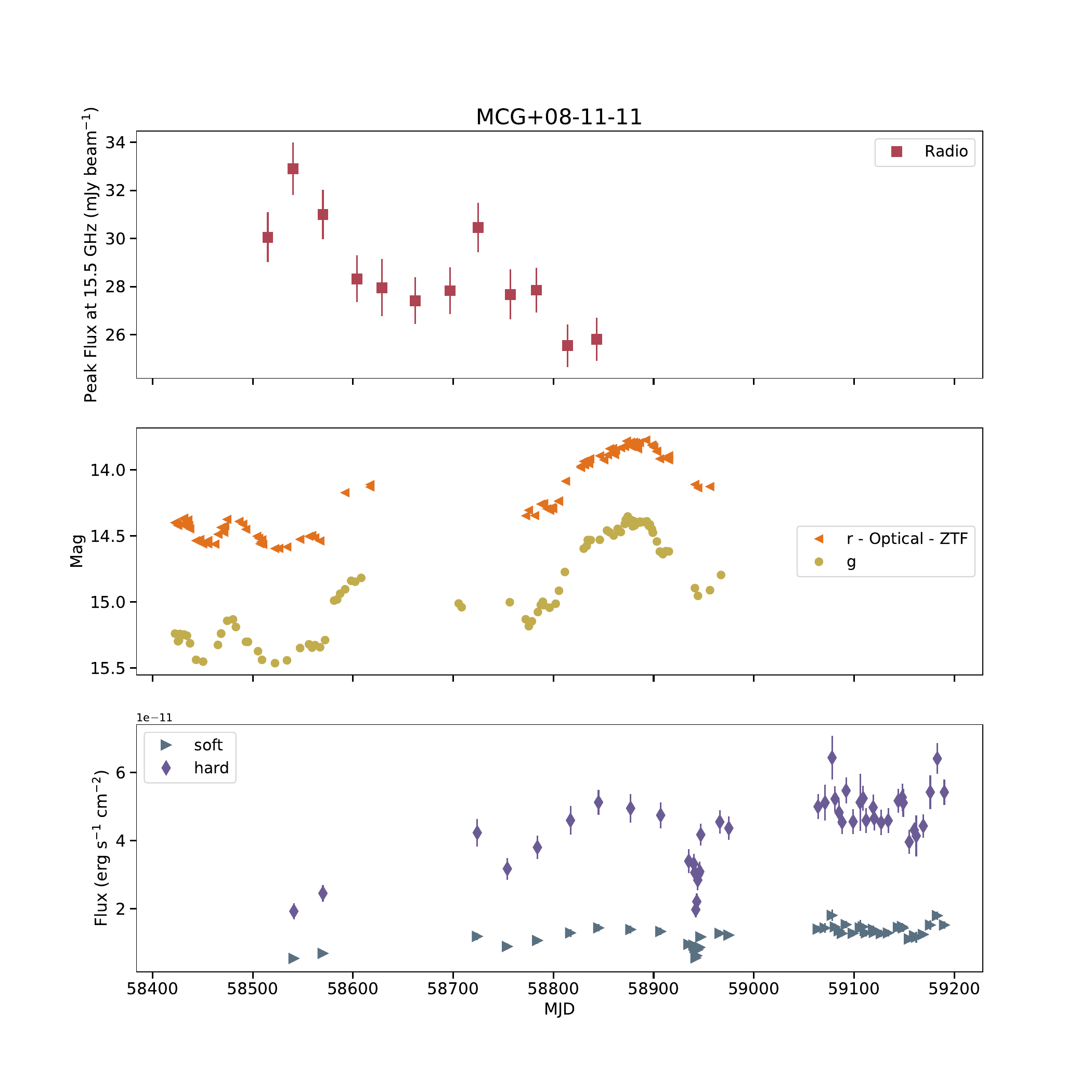}%
    \raisebox{1.5cm}{\includegraphics[width=0.42\textwidth]{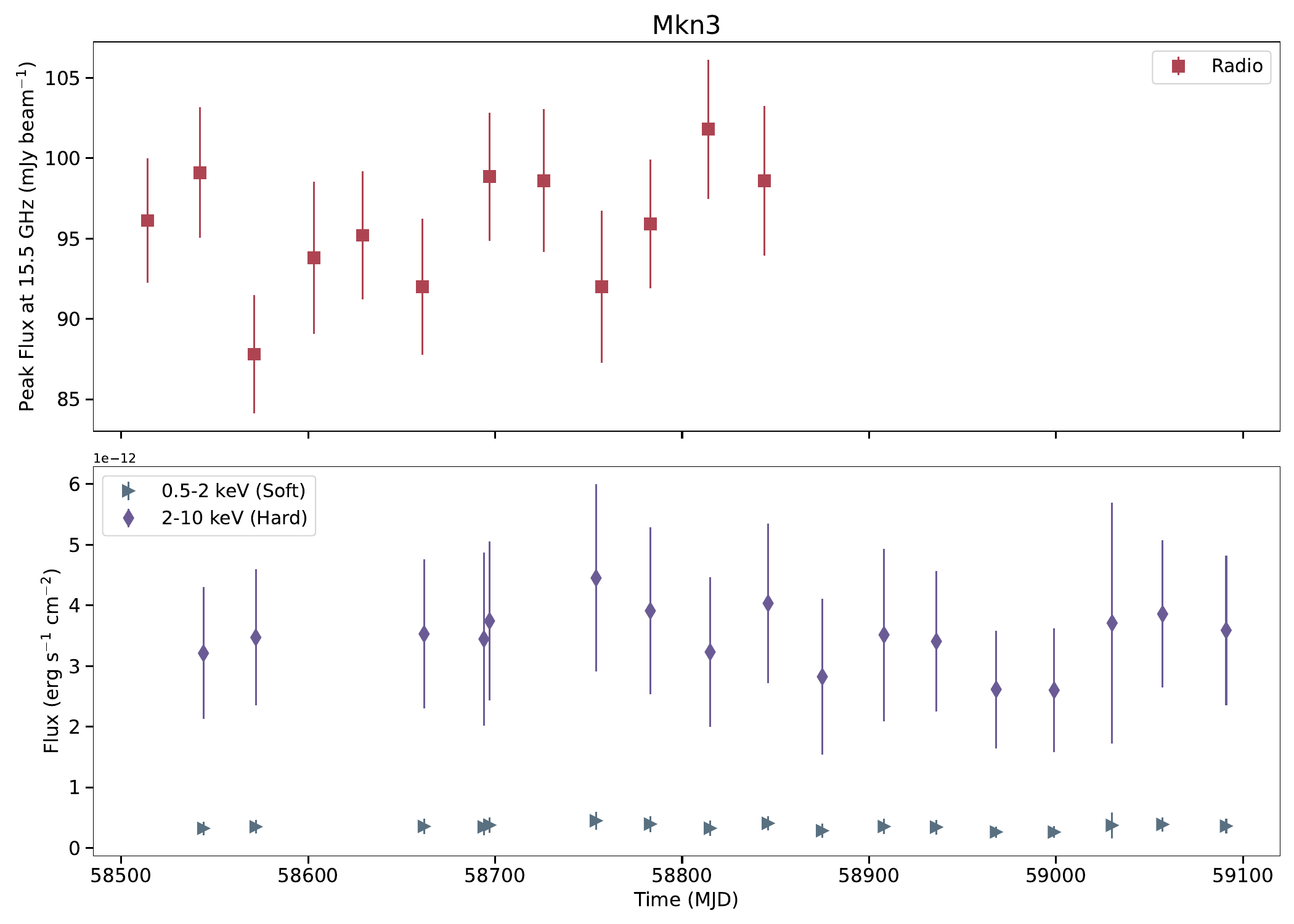}}
    \vspace{-1cm}
    
    \caption{Multiwavelength light curves for the sample. Top panel: Radio flux density at 15~GHz (AMI-LA) in mJy. Middle panel: Optical apparent magnitude from ZTF (g and r bands). Bottom panel: X-ray flux in the soft and hard bands from \textit{Swift}-XRT. Time is reported in MJD and error bars indicate 1$\sigma$ uncertainties. Source names are indicated at the top of each individual panel.}
    \label{fig:lcs_full}
\end{figure*}

\newpage

\begin{figure*}
    \ContinuedFloat
    \centering
    \vspace{-0.8cm}
    \includegraphics[width=0.5\textwidth]{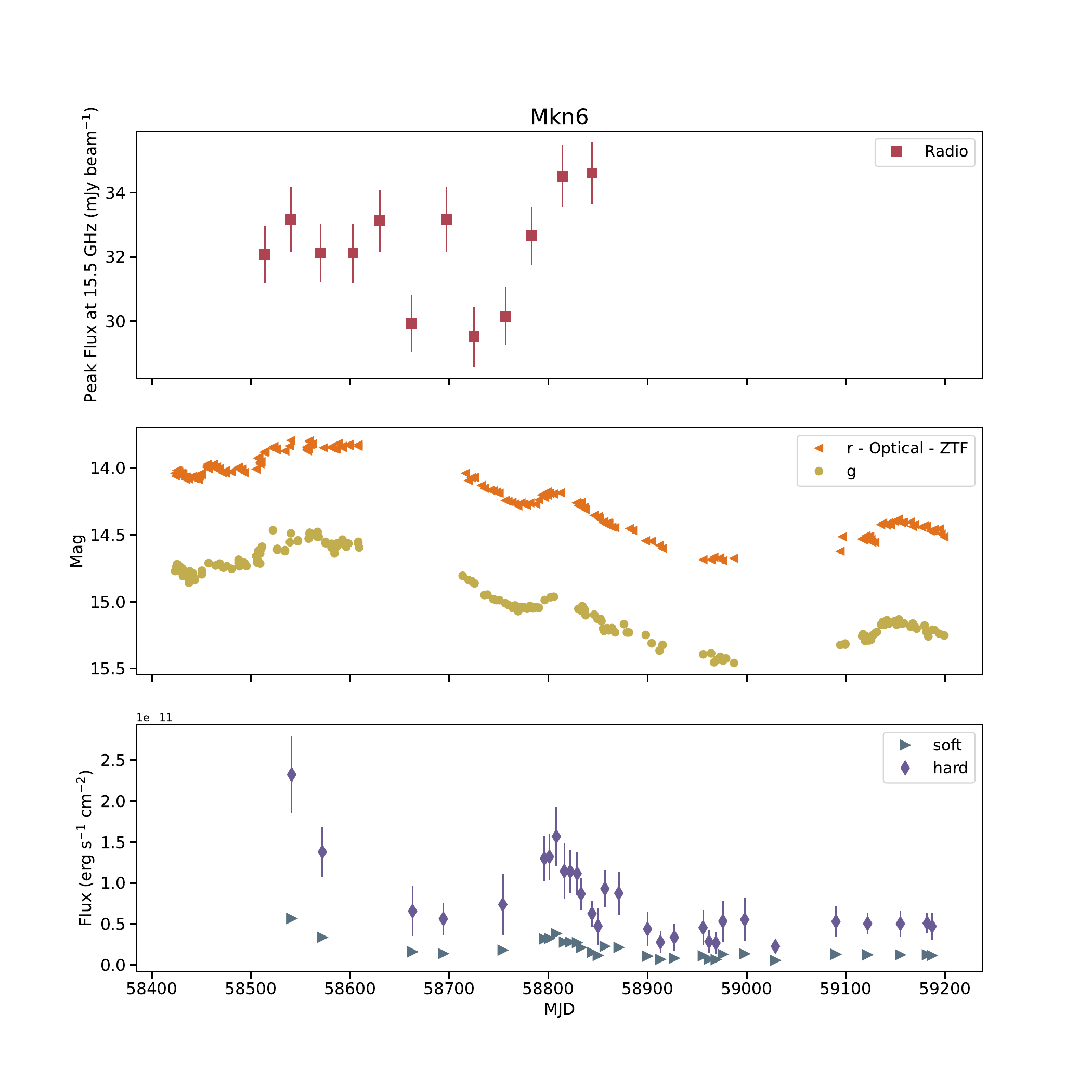}%
    \includegraphics[width=0.5\textwidth]{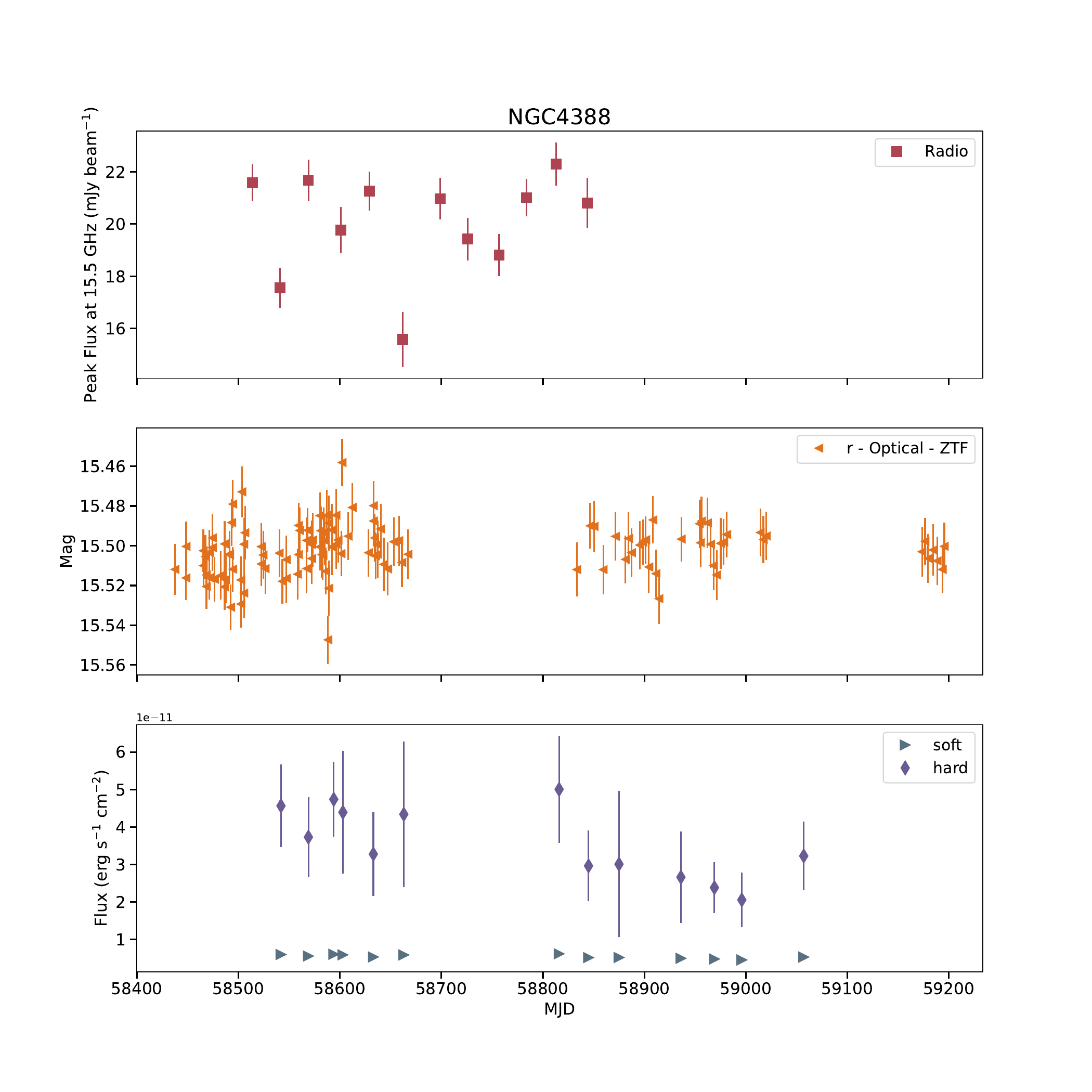}
    \vspace{-1.0cm}
    
    \includegraphics[width=0.5\textwidth]{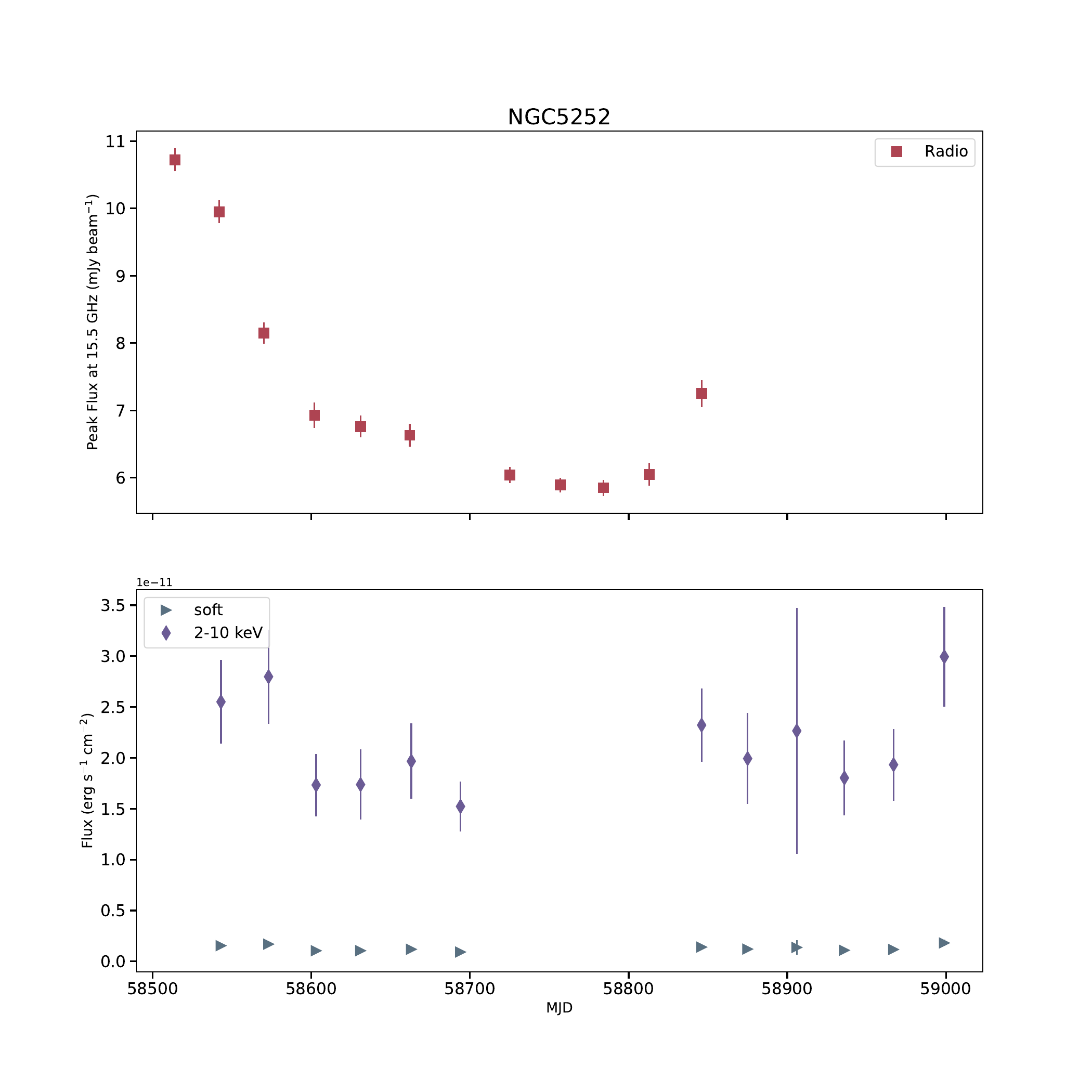}%
    \includegraphics[width=0.5\textwidth]{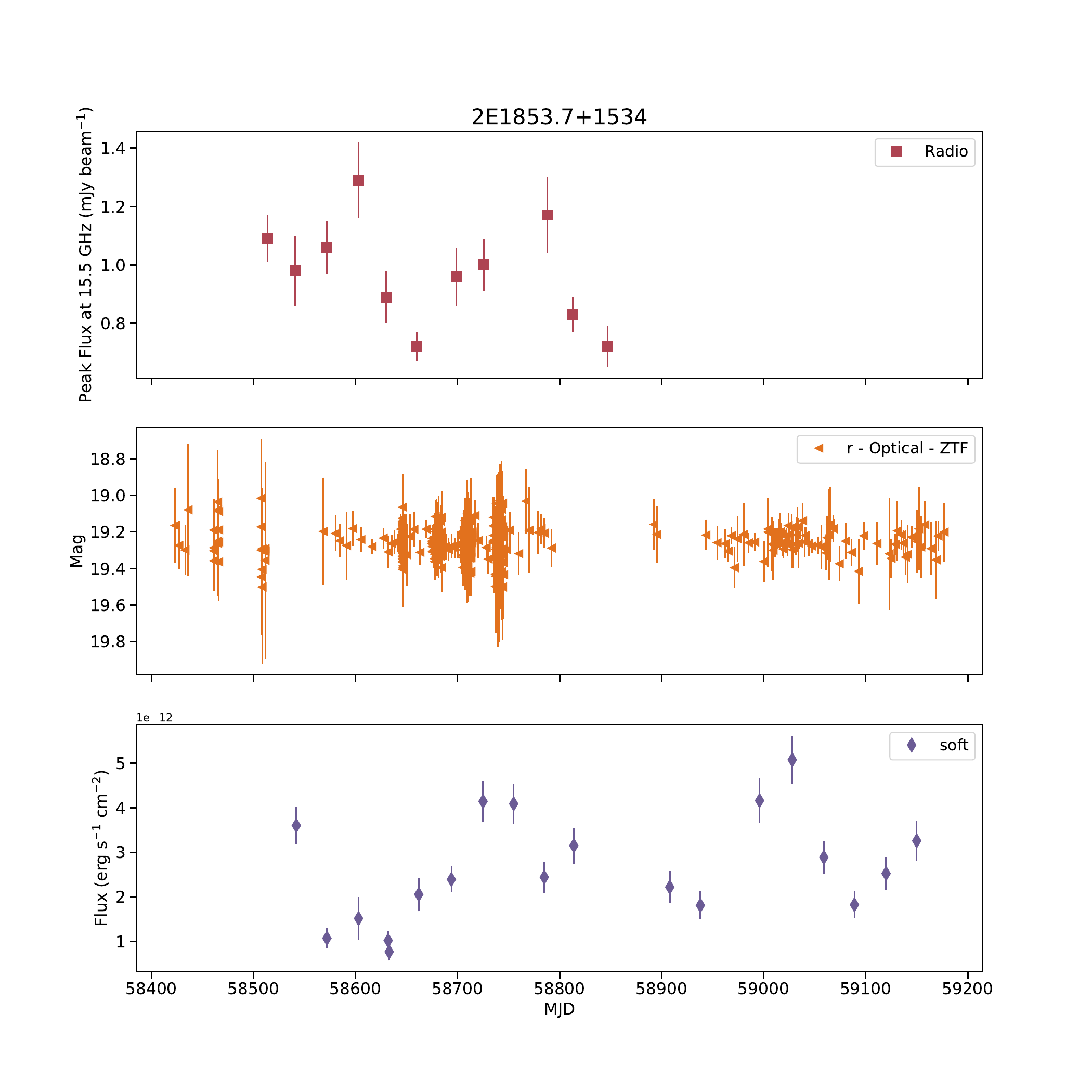}
    \vspace{-1.0cm}
    
    \includegraphics[width=0.5\textwidth]{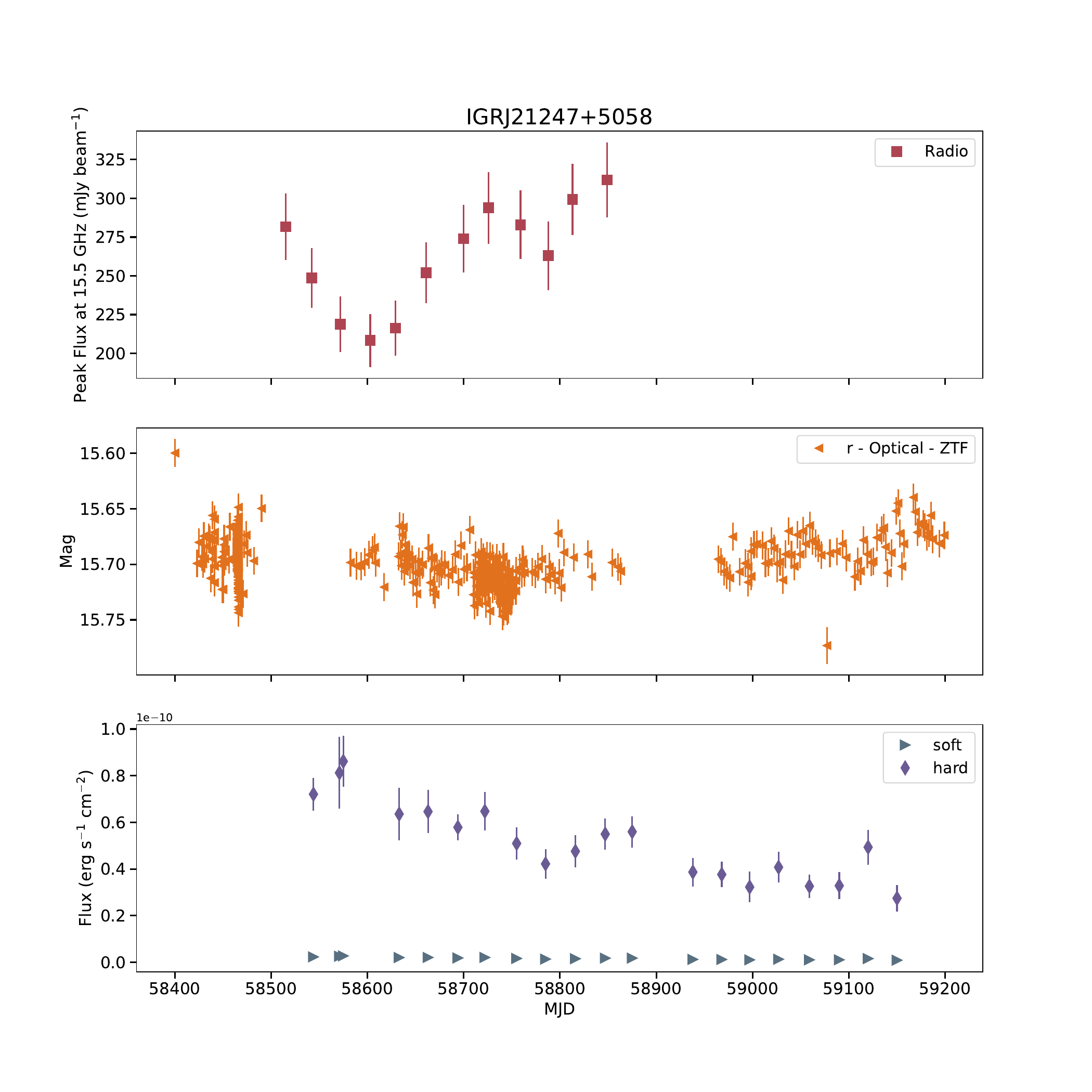}%
    \includegraphics[width=0.5\textwidth]{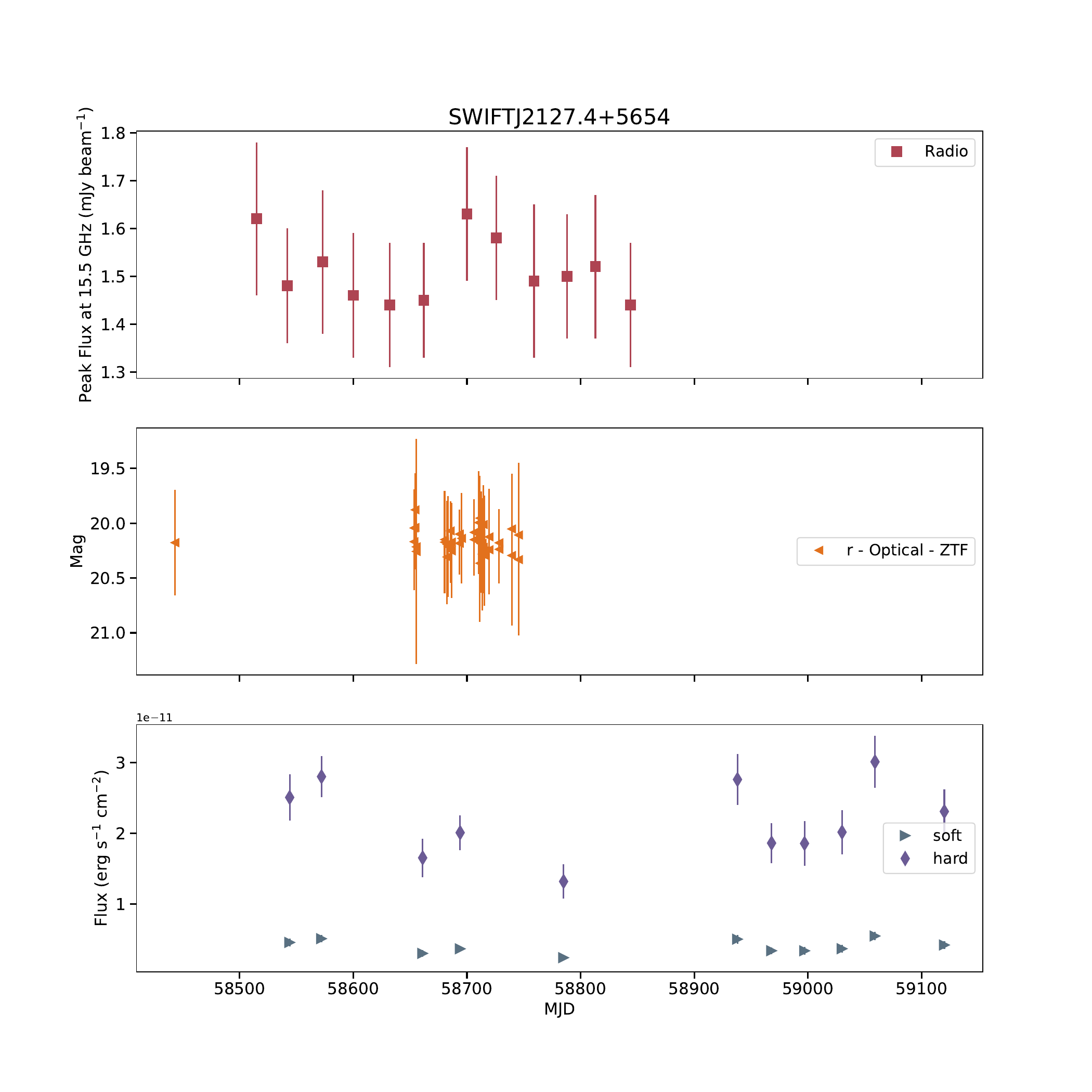}
    
    \caption{(Cont.)}
\end{figure*}

\newpage

\begin{figure*}
    \ContinuedFloat
    \centering
    \includegraphics[width=0.5\textwidth]{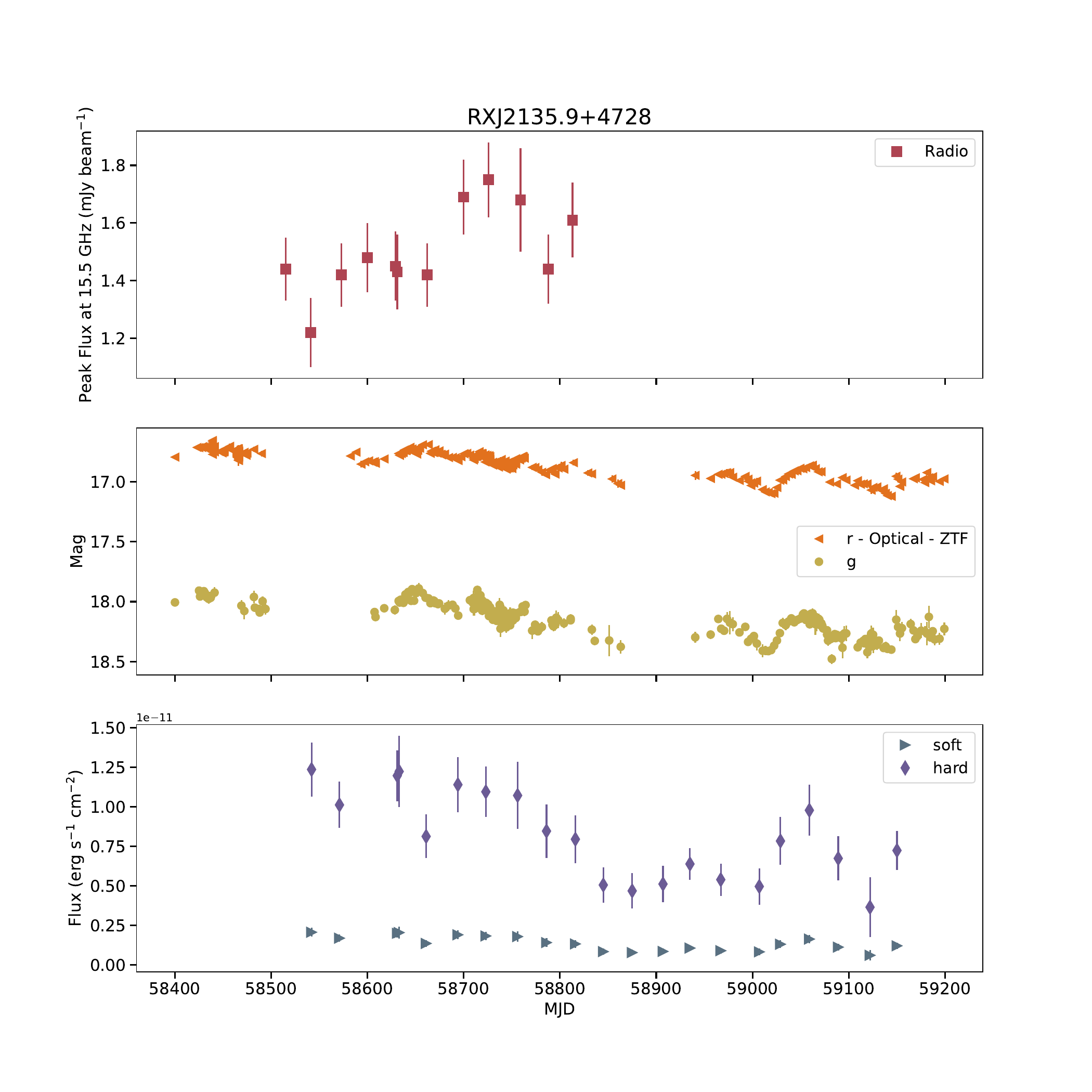}%
    \includegraphics[width=0.5\textwidth]{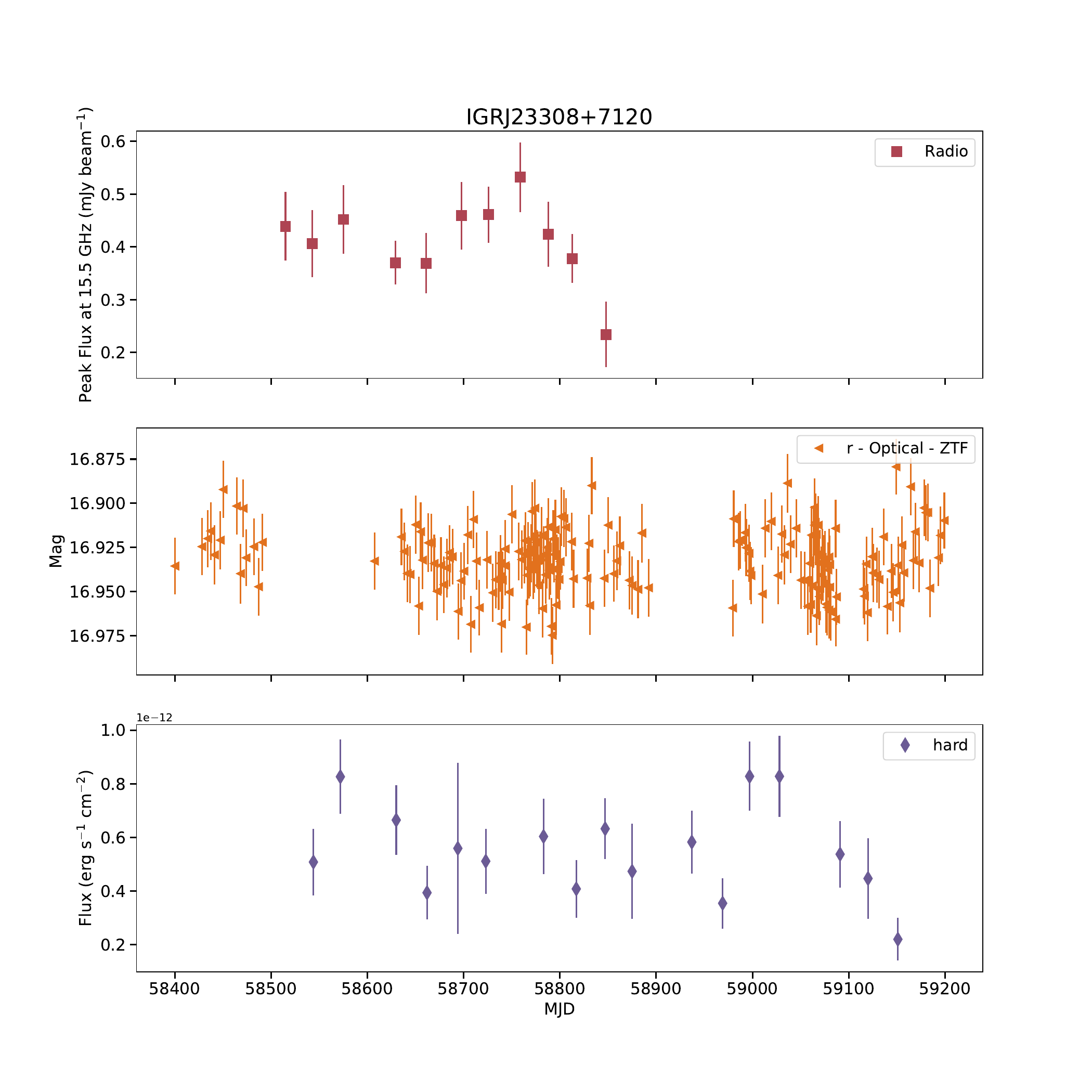} 
    
    \caption{(Cont.)}
\end{figure*}

\newpage

\section{Features}
\label{appendix:features}

This section summarizes the variability features calculated for our sample across the radio, optical, and X-ray bands. Detailed definitions of these features are provided in Section~\ref{sec:analysis}. All reported values and their corresponding uncertainties are listed in Table~\ref{tab:variability_results}.

\onecolumn 
\clearpage

\begin{small}
\setstretch{1.0} 
\begin{longtable}{llccccccc}
\caption{Variability features for the sample. Columns include the reduced chi-squared, normalized excess variance, $\sigma_{NXS}^2$, scaled by $10^{-2}$, fractional variability amplitude, and the Mexican Hat over 70 and 200-day windows. $\Delta (\%)$ refers to the relative difference between the minimum and maximum counts. A dash (--) indicates insufficient data for error estimation.} \label{tab:variability_results} \\
\hline \hline
\toprule
Source & Band & $\chi^2/\mathrm{dof}$ & $\sigma_{nxs}^2$ & $F_{var}$ & MH (70d) & MH (200d) & $\Delta (\%)$ \\ \midrule
\endfirsthead

\multicolumn{8}{c}{{\bfseries \tablename\ \thetable{} -- Continued from previous page}} \\
\toprule
Source & Band & $\chi^2/\mathrm{dof}$ & $\sigma_{nxs}^2$ & $F_{var}$ & MH (70d) & MH (200d) & $\Delta (\%)$ \\ \midrule
\endhead

\midrule
\multicolumn{8}{r}{{Continued on next page}} \\
\bottomrule
\endfoot

\bottomrule
\endlastfoot
\hline
IGR\,J00333+6122  & radio & 39.35/11 & $1.48 \pm 0.63$ & $0.121 \pm 0.003$ & $0.400 \pm 0.226$ & $0.531 \pm 0.344$ & 38.3 \\
& ZTF$_r$ & 328.10/569 & $<0.090$ & $<0$ & $-0.065 \pm 0.004$ & $-0.073 \pm 0.006$ & 80.1 \\
 & hard & 182.81/20 & $28.37 \pm 7.63$ & $0.533 \pm 0.038$ & $2.806 \pm 2.015$ & $4.181 \pm 3.314$ & 94.4 \\
\midrule
QSO\,B0241+62  & radio & 154.35/10 & $5.25 \pm 1.02$ & $0.229 \pm 0.005$ & $0.387 \pm 0.206$ & $0.169 \pm 0.309$ & 52.6 \\
& ZTF$_g$ & 25568.07/220 & $1.60 \pm 0.03$ & $0.127 \pm 0.001$ & $0.040 \pm 0.018$ & $0.342 \pm 0.147$ & 42.8 \\
 & ZTF$_r$ & 46658.11/393 & $0.71 \pm 0.01$ & $0.084 \pm 0.001$ & $0.015 \pm 0.008$ & $0.144 \pm 0.052$ & 34.4 \\
 & hard & 81.18/17 & $6.31 \pm 1.57$ & $0.251 \pm 0.008$ & $0.661 \pm 0.509$ & $3.664 \pm 1.822$ & 73.6 \\
 & soft & 1143.50/17 & $7.77 \pm 0.44$ & $0.279 \pm 0.002$ & $0.676 \pm 0.509$ & $3.680 \pm 1.822$ & 73.6 \\
\midrule
LEDA168563  & radio & 54.54/11 & $0.93 \pm 0.30$ & $0.097 \pm 0.001$ & $0.099 \pm 0.106$ & $0.412 \pm 0.248$ & 29.2 \\
& ZTF$_g$ & 13539.26/166 & $1.37 \pm 0.03$ & $0.117 \pm 0.001$ & $0.026 \pm 0.011$ & $0.196 \pm 0.087$ & 36.4 \\
 & ZTF$_r$ & 28571.89/182 & $1.38 \pm 0.02$ & $0.117 \pm 0.001$ & $0.020 \pm 0.010$ & $0.177 \pm 0.073$ & 34.0 \\
 & hard & 42.49/11 & $26.95 \pm 8.15$ & $0.519 \pm 0.041$ & $3.134 \pm 2.687$ & $4.014 \pm 2.690$ & 78.0 \\
 & soft & 345.47/11 & $32.72 \pm 3.02$ & $0.572 \pm 0.015$ & $3.198 \pm 2.687$ & $4.078 \pm 2.690$ & 78.0 \\
\midrule
4U\,0517+17  & radio & 23.40/11 & $0.20 \pm 0.16$ & $0.044 \pm 0.001$ & $0.103 \pm 0.077$ & $0.193 \pm 0.107$ & 18.5 \\
& ZTF$_g$ & 71327.09/134 & $7.60 \pm 0.07$ & $0.276 \pm 0.001$ & $0.081 \pm 0.020$ & $0.086 \pm 0.116$ & 60.3 \\
 & ZTF$_r$ & 74947.76/142 & $5.64 \pm 0.05$ & $0.237 \pm 0.001$ & $0.065 \pm 0.017$ & $0.095 \pm 0.095$ & 57.0 \\
 & hard & 207.87/11 & $16.49 \pm 3.48$ & $0.406 \pm 0.017$ & $0.514 \pm 0.400$ & $1.327 \pm 1.546$ & 79.6 \\
 & soft & 1071.90/11 & $18.18 \pm 1.53$ & $0.426 \pm 0.008$ & $0.532 \pm 0.400$ & $1.346 \pm 1.546$ & 79.6 \\
\midrule
MCG+08-11-11  & radio & 51.13/11 & $0.44 \pm 0.14$ & $0.067 \pm 0.001$ & $0.056 \pm 0.038$ & $0.118 \pm 0.081$ & 22.4 \\
& ZTF$_g$ & 449625.21/86 & $11.18 \pm 0.05$ & $0.334 \pm 0.001$ & $0.102 \pm 0.054$ & $2.929 \pm 1.346$ & 64.0 \\
 & ZTF$_r$ & 236902.86/99 & $6.96 \pm 0.04$ & $0.264 \pm 0.001$ & $0.042 \pm 0.027$ & $1.383 \pm 0.597$ & 53.0 \\
 & hard & 557.11/44 & $5.36 \pm 0.60$ & $0.232 \pm 0.003$ & $0.243 \pm 0.094$ & $1.567 \pm 0.723$ & 70.1 \\
 & soft & 1452.45/44 & $5.81 \pm 0.38$ & $0.241 \pm 0.002$ & $0.248 \pm 0.094$ & $1.572 \pm 0.723$ & 70.1 \\
\midrule
Mkn\,3  & radio & 10.32/11 & $<0.090$ & $<0$ & $0.056 \pm 0.036$ & $0.053 \pm 0.034$ & 13.8 \\
& hard & 2.75/16 & $<0.01$ & -- & -- & $0.607 \pm 0.381$ & 41.6 \\
 & soft & 3.83/16 & -- & -- & -- & $0.638 \pm 0.381$ & 41.6 \\
\midrule
Mkn\,6  & radio & 34.73/11 & $0.18 \pm 0.08$ & $0.043 \pm 0.001$ & $0.073 \pm 0.043$ & $0.069 \pm 0.051$ & 14.7 \\
& ZTF$_g$ & 350673.77/174 & $6.59 \pm 0.03$ & $0.257 \pm 0.001$ & $0.037 \pm 0.013$ & $0.162 \pm 0.165$ & 59.9 \\
 & ZTF$_r$ & 145373.67/179 & $4.96 \pm 0.03$ & $0.223 \pm 0.001$ & $0.032 \pm 0.012$ & $0.175 \pm 0.136$ & 56.3 \\
 & hard & 188.79/30 & $36.10 \pm 6.62$ & $0.601 \pm 0.033$ & $1.480 \pm 0.554$ & $9.074 \pm 6.817$ & 90.2 \\
 & soft & 6911.37/30 & $44.30 \pm 1.16$ & $0.666 \pm 0.006$ & $1.576 \pm 0.554$ & $9.169 \pm 6.817$ & 90.2 \\
\midrule
NGC\,4388  & radio & 53.92/11 & $0.78 \pm 0.22$ & $0.088 \pm 0.001$ & $0.424 \pm 0.253$ & $0.232 \pm 0.179$ & 30.1 \\
& ZTF$_r$ & 122.52/118 & $<0.001$ & $<0$ & $0.000 \pm 0.000$ & $0.000 \pm 0.000$ & 7.9 \\
 & hard & 78.94/12 & $4.12 \pm 1.42$ & $0.203 \pm 0.007$ & $1.189 \pm 0.349$ & $0.822 \pm 0.474$ & 52.8 \\
 & soft & 32.49/12 & $5.49 \pm 0.01$ & $0.234 \pm 0.001$ & $1.205 \pm 0.349$ & $0.837 \pm 0.474$ & 52.8 \\
\midrule
NGC\,5252  & radio & 1179.03/10 & $5.14 \pm 0.30$ & $0.227 \pm 0.001$ & $0.066 \pm 0.108$ & $0.826 \pm 0.608$ & 45.4 \\
& hard & 22.59/11 & $1.25 \pm 1.80$ & $0.112 \pm 0.009$ & $0.598 \pm 0.489$ & $1.413 \pm 0.747$ & 49.1 \\
 & soft & 1301.67/11 & $4.52 \pm 0.30$ & $0.213 \pm 0.001$ & $0.640 \pm 0.489$ & $1.455 \pm 0.747$ & 49.1 \\
\midrule
2E\,1853.7+1534  & radio & 56.80/10 & $2.45 \pm 0.97$ & $0.157 \pm 0.005$ & $0.746 \pm 0.508$ & $1.310 \pm 0.726$ & 44.2 \\
& ZTF$_r$ & 164.57/473 & $<0.01$ & -- & $-0.015 \pm 0.001$ & $-0.004 \pm 0.005$ & 36.1 \\
 & hard & 273.22/18 & $18.90 \pm 2.90$ & $0.435 \pm 0.015$ & $1.252 \pm 1.064$ & $12.330 \pm 5.685$ & 84.9 \\
\midrule
IGR\,J21247+5058  & radio & 34.05/11 & $1.09 \pm 0.54$ & $0.104 \pm 0.003$ & $0.067 \pm 0.069$ & $0.421 \pm 0.279$ & 33.2 \\
& ZTF$_r$ & 1322.52/757 & $0.01 \pm 0.00$ & $0.010 \pm 0.000$ & $0.001 \pm 0.000$ & $0.001 \pm 0.001$ & 14.7 \\
 & hard & 118.56/19 & $8.73 \pm 1.83$ & $0.295 \pm 0.009$ & $0.221 \pm 0.186$ & $1.063 \pm 0.617$ & 68.2 \\
 & soft & 3313.57/19 & $10.41 \pm 0.36$ & $0.323 \pm 0.002$ & $0.239 \pm 0.186$ & $1.082 \pm 0.617$ & 68.2 \\
\midrule
SWIFT\,J2127.4+5654  & radio & 2.59/11 & $<0.01$ & -- & $0.021 \pm 0.025$ & $0.080 \pm 0.053$ & 11.7 \\
& ZTF$_r$ & 2.47/40 & $<0.01$ & -- & $-0.137 \pm 0.012$ & $-0.153 \pm 0.014$ & 36.2 \\
 & hard & 39.54/10 & $4.28 \pm 1.72$ & $0.207 \pm 0.009$ & $0.311 \pm 0.272$ & $2.069 \pm 1.254$ & 56.2 \\
 & soft & 863.59/10 & $5.81 \pm 0.39$ & $0.241 \pm 0.002$ & $0.327 \pm 0.272$ & $2.084 \pm 1.254$ & 56.2 \\
\midrule
RX\,J2135.9+4728  & radio & 15.16/11 & $0.30 \pm 0.39$ & $0.055 \pm 0.002$ & $0.096 \pm 0.075$ & $0.339 \pm 0.261$ & 30.3 \\
& ZTF$_g$ & 6306.74/289 & $1.23 \pm 0.04$ & $0.111 \pm 0.001$ & $0.037 \pm 0.009$ & $0.196 \pm 0.074$ & 41.8 \\
 & ZTF$_r$ & 31545.43/765 & $0.38 \pm 0.01$ & $0.062 \pm 0.000$ & $0.007 \pm 0.002$ & $0.033 \pm 0.012$ & 34.8 \\
 & hard & 85.62/20 & $8.76 \pm 2.39$ & $0.296 \pm 0.012$ & $0.556 \pm 0.402$ & $1.429 \pm 0.916$ & 70.5 \\
 & soft & 990.39/20 & $11.45 \pm 0.74$ & $0.338 \pm 0.004$ & $0.585 \pm 0.402$ & $1.458 \pm 0.916$ & 70.5 \\
\midrule
IGR\,J23308+7120  & radio & 15.65/10 & $1.39 \pm 1.34$ & $0.118 \pm 0.007$ & $0.258 \pm 0.290$ & $1.531 \pm 0.932$ & 56.0 \\
& ZTF$_r$ & 224.49/190 & $<0.001$ & $<0$ & $0.000 \pm 0.000$ & $0.001 \pm 0.001$ & 8.4 \\
 & hard & 40.28/16 & $3.47 \pm 3.06$ & $0.186 \pm 0.015$ & $1.771 \pm 1.189$ & $3.312 \pm 1.770$ & 73.4 \\
 \hline
\end{longtable}
\end{small}
\twocolumn %

\label{lastpage}
\end{document}